\documentclass[aps,superscriptaddress,nofootinbib,showpacs,eqsecnum,preprint,tightenlines]{revtex4}
\usepackage{hyperref}
\usepackage{epsfig,rotating}
\usepackage{amsmath,amssymb}
\usepackage{dsfont}
\usepackage{bbm}
\usepackage{slashed}
\numberwithin{equation}{section}

\newcommand{\be}{\begin{equation}}
\newcommand{\ee}{\end{equation}}
\newcommand{\bea}{\begin{eqnarray}}
\newcommand{\eea}{\end{eqnarray}}

\newcommand{\vx}{\vec{x}}

\newcommand{\vp}{\vec{p}}

\newcommand{\vk}{\vec{k}}

\begin{document}

\title{Pre-slow roll initial conditions: large scale power suppression and  infrared aspects during inflation.}

\author{Louis Lello}
\email{lal81@pitt.edu}

\author{Daniel Boyanovsky}
\email{boyan@pitt.edu} \affiliation{Department of Physics and
Astronomy, University of Pittsburgh, Pittsburgh, PA 15260}

\author{Richard Holman}
\email{rh4a@andrew.cmu.edu} \affiliation{Department of Physics, \\
Carnegie Mellon University, Pittsburgh PA 15213}

\date{\today}

\begin{abstract}
 If the large scale anomalies in the temperature power spectrum of the cosmic microwave background  are of primordial origin, they may herald modifications to the slow roll inflationary paradigm on the largest scales. We study the possibility that the origin of the large scale power suppression is a modification of initial conditions during slow roll as a result of a pre-slow roll phase during which the inflaton evolves rapidly. This stage is manifest in a potential in the equations for the Gaussian fluctuations during slow roll and modify the power spectra of scalar   perturbations via an initial condition transfer function $\mathcal{T}(k)$. We provide a general   analytical   study of its large and small scale properties and analyze the impact of these initial conditions on the infrared aspects of typical test scalar fields. The infrared behavior of massless minimally coupled test scalar field theories leads to the dynamical generation of mass and anomalous dimensions, both depend non-analytically on $\mathcal{T}(0)$. During inflation all quanta decay into many quanta even of the same field because of the lack of kinematic thresholds. The decay leads to a quantum entangled state of sub and superhorizon quanta with correlations across the horizon.  We find the modifications of the decay width and the entanglement entropy from the initial conditions. In all cases, initial conditions from a ``fast-roll'' stage that lead to a suppression in the scalar power spectrum at large scales also result in a suppression of the dynamically generated masses, anomalous dimensions and decay widths.

\end{abstract}


\maketitle

\section{Introduction}
 Inflation   provides a solution to the horizon and flatness problems and a mechanism for generating scalar (curvature) and tensor (gravitational wave)   quantum fluctuations\cite{staro2,guth,linde,al}  which seed the small temperature inhomogeneities in the CMB upon reentering the particle horizon during recombination.   Although there are several different inflationary scenarios  most of them predict  a  nearly gaussian and nearly scale invariant power spectrum of adiabatic fluctuations.For reviews see\cite{mukh,kolb,riotto2,baumann,giov}.

Observations of the cosmic microwave background (CMB) offer compelling evidence in support of the inflationary paradigm, confirming that anisotropies are well described by adiabatic, gaussian and nearly scale invariant fluctuations\cite{wmap7,wmap9,planck} and are beginning to discriminate among different scenarios.

Recent results from the Planck collaboration\cite{planck} have provided the most precise analysis of the (CMB) to date, confirming the main features of the inflationary paradigm, but at the same time highlighting perplexing large scale anomalies, some of them, such as a low quadrupole,  dating back to the early observations of the Cosmic Background Explorer (COBE)\cite{cobe,bondlow}, confirmed with greater accuracy by WMAP\cite{wmaplow} and Planck\cite{planck}. The most recent Planck\cite{planck} data still finds  a statistically significant discrepancy at low multipoles, reporting a power deficit $5-10\%$ at $l \lesssim 40$ with $2.5-3\,\sigma$ significance. This puzzling and persistent result stands out in an otherwise consistent picture of $\Lambda CDM$ insofar as the (CMB) power spectrum is concerned.

The large scale suppression of the primordial power spectrum on super Hubble scales and its statistical significance was analyzed early on\cite{berera1,berera2} within the context of the original COBE-DMR data. These references report on  a systematic analysis of possible mechanisms for large scale suppression (see refs.\cite{berera1,berera2} and references therein) and their statistical significance with the conclusion that in inflationary scenarios, the suppression on superHubble scales are expected to be of low statistical significance. The latest results from Planck\cite{planck} rule out foreground contamination as the origin of the large scale suppression but also highlights that at $2.5-3\,\sigma$ this suppression is still of low statistical significance, and obviously cosmic variance limited. The conclusion of ref.\cite{berera2} is that if suppression on Hubble scales is indeed measured with sufficient statistical significance, complementary measurements such as polarization for example, can serve as consistency checks.

The interpretation and statistical significance of these anomalies is a matter of much debate, but being associated with the largest scales, hence the most primordial aspects of the power spectrum, their observational evidence is not completely dismissed. The possible origin of the large scale anomalies is vigorously discussed, whether these are of primordial origin or a consequence of the statistical analysis (masking) or secondary anisotropies is still an open question. Recent studies claim the removal of some of the large scale anomalies (including the suppression of power of the low multipoles) after substraction of the integrated Sachs-Wolfe effect\cite{francis,rassat}, however a different   analysis of the WMAP9\cite{wmap9} data still finds a statistically significant discrepancy at low multipoles\cite{grup}, suggesting that the possibility  of the primordial origin of the large scale anomalies   merits further study. Recent analysis of this lack of power at low $l$\cite{grup} and large angles\cite{copi}, suggests that while limited by cosmic variance,  the possibility  of the primordial origin of the large scale anomalies cannot be dismissed and  merits further study.

The simpler inflationary paradigm that successfully explains the cosmological data relies on the dynamics of a scalar field, the inflaton, evolving slowly during the inflationary stage with the dynamics determined by a fairly flat potential. This simple, yet observationally supported inflationary scenario is referred to as slow-roll inflation\cite{kolb,mukh,riotto2,baumann,giov}. Within this   scenario   wave vectors of cosmological relevance    cross the Hubble radius during inflation with nearly constant amplitude leading to a nearly scale invariant power spectrum. The quantization of the gaussian fluctuations (curvature and tensor) is carried out by imposing a set of initial conditions so that fluctuations with wavevectors deep inside the Hubble radius are described by Minkowski space-time free field mode functions. These are known as Bunch-Davies initial conditions\cite{bunch} (see for example\cite{kolb,mukh,baumann,giov} and references therein).

The issue of modifications of these initial conditions and the potential impact on the inflationary power spectra\cite{ini1,ini2,ini3,ini4,ini5,holini1,holini2,martin,daniels,picon,berera1,berera2},  enhancements to non-gaussianity\cite{holtol1,holtol2,ganc,parker1,parker2,porto,dustin}, and   large scale structure\cite{ganckoma} have been discussed in the literature. Whereas the recent results from Planck\cite{planck} provide tight constraints on primordial non-gaussianities including modifications from initial conditions, these constraints \emph{per se} do not apply directly to the issue of initial conditions on other observational aspects.

Non-Bunch-Davies initial conditions arising from a pre-slow roll stage during which the (single) inflaton field features a ``fast-roll'' dynamics have been proposed as a \emph{possible} explanation of power suppression at large scales\cite{lindefast,contaldi,boyan3,hectordestri,reviunos,lasenby}. More recently a detailed analysis of modifications of power spectra for curvature and tensor perturbations from a kinetic dominated pre-slow roll stage has been reported\cite{lellor}.

 Alternative pre-slow-roll descriptions in terms of interpolating scale factors pre (and post) inflation have been discussed in ref.\cite{parkglenz} and the impact of initial conditions from high energy models on power spectra and non-gaussianities and the tensor to scalar ratio was studied in ref.\cite{schwarz,amjad,kundu,jain}.

 The largest scales that manifest the suppression in the power spectrum correspond to fluctuations whose wavevectors exited the Hubble radius about 60-e-folds before the end of inflation, therefore \emph{if} the large scale anomalies are of primordial origin and herald new physics, an explanation must be sought in the
\emph{infrared} sector of inflationary perturbations.

 It has been recognized that the contribution from super-Hubble fluctuations of  massless (or nearly massless) fields in de Sitter (or nearly de Sitter) inflation to loop corrections of cosmological correlation functions lead to infrared and secular divergences that hinder the reliability of the perturbative expansion\cite{weinberg,seery,branrecent,giddins,hebe,bran,woodard,rajaraman}. These divergences    invalidate the semiclassical approximation\cite{holmanburgess} and require  non-perturbative resummations\cite{riottosloth,enq,boyholds,boyquasi,serreau1} or kinetic\cite{akhmedov} treatments.

In the seminal work of ref.\cite{staro1} it was shown that resummation of infrared and secular divergences leads to the dynamical generation of mass, a result that was further explored in ref.\cite{richard} and more recently  a self-consistent mechanism of  mass generation for scalar fields through infrared fluctuations has been suggested\cite{holmanburgess,rigo,garb,arai,rajaraman,serreau2,boyquasi}.

Furthermore,  the lack of a global time-like killing vector leads to remarkable physical effects in inflationary cosmology, for example it implies the lack of kinematic thresholds (a direct consequence of energy-momentum conservation) and the decay of   fields even in their own quanta\cite{boyprem,boyan,boyquasi} a result that was  also   investigated  for massive fields in ref.\cite{moschella,leblond}  and confirmed in general  in ref.\cite{marolf}.

If a parent particle decays into two or more daughter particles, the quantum state that describes the daughter particles is an \emph{entangled state}\cite{lellomink}, the entanglement is a consequence of conservation laws, such as momentum, angular momentum etc.
Recently it was recognized that in inflationary cosmology the decay of a particle into sub and superhorizon quanta produces an \emph{entangled state} with quantum correlations across the Hubble radius\cite{lellosup}.

\vspace{2mm}

\textbf{Motivations, goals and summary of results:} Our study is motivated by the persistency of the power suppression at large scales as evidenced in the Planck data\cite{planck} and the possibility that these anomalies are of primordial origin and reflect novel physical infrared effects with observational consequences.

Our goals in this article are two-fold: i) to study in detail the modifications of  initial conditions within the paradigm of single field inflation but described by an early, pre-slow-roll stage in which the inflaton field undergoes ``fast-roll'' dynamics as proposed in refs.\cite{lindefast,contaldi,boyan3,hectordestri,reviunos,lasenby} and more recently in ref.\cite{lellor}, ii) to assess how the modified initial conditions impact infrared phenomena in scalar fields. In particular we focus on the impact of non-Bunch-Davies initial conditions as a consequence of a pre-slow-roll   stage on non-perturbative phenomena, such as dynamical mass generation, decay of quanta and superhorizon correlations arising from the quantum entanglement of the daughter particles. To the best of our knowledge, the effect of initial conditions on infrared effects such as dynamical mass generation and decay of single particle excitations has not been studied.

We consider the case in which non-Bunch-Davies initial conditions during inflation are a consequence of a pre-slow roll stage during which the inflaton undergoes fast roll dynamics. This ``fast-roll'' stage prior to slow roll  results in a potential in the equations of motion of gaussian fluctuations and lead to a change of initial conditions during the slow roll stage via Bogoliubov coefficients.

 We begin with a description of a fast-roll stage dominated by the kinetic term of the inflaton and follow with a detailed analysis of superhorizon and subhorizon behavior of the Bogoliubov coefficients describing non-Bunch-Davies initial conditions during the inflationary stage and how these modify the large scale power spectrum of fluctuations.  The effect of these non-Bunch-Davies initial conditions is encoded in the power spectra of scalar   perturbations via a \emph{transfer function} $\mathcal{T}(k)$.

  Implementing methods from the quantum theory of scattering, we provide a general  analytic study of the superhorizon and subhorizon limits of the initial condition transfer function $\mathcal{T}(k)$ and find that for superhorizon momenta
  \[\mathcal{T}(k) \simeq \mathcal{T}(0) +\mathcal{O}(k^2)\]
  and obtain an explicit expression for $\mathcal{T}(0)$  For subhorizon momenta we find
  \[\mathcal{T}(k) \simeq 1 +\mathcal{O}(1/k^4)\,.\]

  We extract the form of the mode functions modified by these initial conditions  in the superhorizon limit and study in detail how this transfer function modifies the infrared behavior in typical scalar field theories, in particular the modification of dynamically generated masses and the width of the single particle states.

  We find that the dynamically generated masses induced by these infrared divergences depend non-analytically on the transfer function. As a consequence of dynamical mass generation the scalar power spectrum features anomalous dimensions that depend non-analytically on $\mathcal{T}(0)$. The decay width of single particle quanta and the entanglement entropy from integrating out superhorizon fluctuations depend also on this quantity.

  We find that a kinetic dominated fast roll stage prior to slow roll leads to an \emph{attractive} potential in the scalar mode equations leading to  $|\mathcal{T}(0)|<1$ and the power spectrum and infrarred correlators are suppressed at large scales. This suppression is also manifest in the dynamically generated masses, anomalous dimensions, decay widths and entanglement entropy.

  \section{Fast roll stage:}

  In this section we summarize the main aspects of a kinetic dominated pre-slow roll stage or ``fast-roll'' stage. More details and a complete analysis of the matching to slow roll can be found in ref.\cite{lellor}.

 We consider a spatially flat Friedmann-Robertson-Walker (FRW) cosmology with
\be ds^2 = dt^2-a^2(t)(d\vec x)^2 =
C^2(\eta)[d\eta^2 - (d\vec x)^2]~~;~~C(\eta) \equiv a(t(\eta)) \; , \ee where $t$ and $\eta$ stand
for cosmic and conformal time respectively. The dynamics of the scale factor in single field inflation is determined by Friedmann and covariant conservation equations
\be H^2 = \frac{1}{3M^2_{Pl}}\Bigg[\frac{1}{2} \dot{\Phi}^2 + V(\Phi)\Bigg]~~;~~ \ddot{\Phi}+3H\dot{\Phi}+V'(\Phi) = 0\,. \label{inflaton}\ee
During the slow roll near de Sitter stage,
\be H^2_{sr} \simeq  \frac{V_{sr}(\Phi)}{3M^2_{Pl}} ~~;~~  3H\dot{\Phi}+V'_{sr}(\Phi) \simeq  0\,. \label{slowly}\ee This stage is   characterized by the smallness of the (potential) slow roll parameters\cite{mukh,kolb,riotto2,baumann,giov}
\be \epsilon_V = \frac{M^2_{Pl}}2 \;
\left[\frac{V'_{sr}(\Phi )}{V_{sr}(\Phi )} \right]^2  \simeq \frac{\dot{\Phi}^2_{sr}}{2M^2_{Pl}H^2}    \quad , \quad
\eta_V = M^2_{Pl} \; \frac{V''_{sr}(\Phi)}{V_{sr}(\Phi)}   \; , \label{slowroll}
\ee (here $M_{Pl}=1/\sqrt{8\pi\,G}$ is the \emph{reduced} Planck mass). The potential slow roll parameters $\epsilon_V,\eta_V$ which have been constrained by Planck and WMAP-polarization (Planck+WP)\cite{planck} to be $\epsilon_V < 0.008 ~ (95\% \,CL);\eta_V = -0.010^{+0.005}_{-0.011}$.

Instead, in this section we consider   an initial stage dominated by the kinetic term, namely a fast roll stage,  thereby neglecting the term $V'$ in the equation of motion for the inflaton,  (\ref{inflaton}) and consider the potential to be (nearly) constant and equal to the potential during the slow roll stage, namely $V(\Phi) \simeq V(\Phi_{sr})\equiv V_{sr}$.
\bea && H^2    =  \Big(\frac{\dot{a}}{a}\Big)^2 = \frac{1}{3M^2_{Pl}}\Bigg[\frac{1}{2} \dot{\Phi}^2 + V_{sr} \Bigg]\label{FRW2}\\ && \ddot{\Phi}+3H\dot{\Phi}  \simeq 0\,. \label{inflaton2}\eea   The solution to (\ref{inflaton2}) is given by
\be \dot{\Phi}(t)= \dot{\Phi}_i \Big(\frac{a_i}{a(t)}\Big)^3 \,, \label{solinfla}\ee an initial value of the velocity damps out and the slow roll stage begins when $\ddot{\Phi} \ll 3H_{sr}\dot{\Phi}\simeq -V'_{sr}(\Phi)$. During the slow roll stage when $3H_{sr}\dot{\Phi}_{sr} \simeq -V'_{sr}$ it follows that
\be   \frac{3\dot{\Phi}^2_{sr}}{2V_{sr} }  = \epsilon_V \,.\label{fidotsr}\ee

The dynamics enters the slow roll stage when $\dot{\Phi} \sim \mathcal{O}(\sqrt{\epsilon_V})$ as seen by (\ref{slowroll}). To a first approximation, we will assume that Eq.(\ref{solinfla}) holds not only for the kinetically dominated epoch, but also until the beginning of slow roll ($\dot{\Phi}^2 \sim \epsilon_V$).   The dynamics enters the slow roll stage at a value of the scale factor $a(t_{sr})\equiv a_{sr}$ so that
\be \dot{\Phi}_{sr} a^3_{sr} = \dot{\Phi}_i a^3_i \,.\label{equ}\ee  We now use the freedom to rescale the scale factor to set
 \be a(t_{sr}) = a_{sr} = 1\,, \label{choice}\ee this normalization is particularly convenient to establish
 when a particular mode crosses the Hubble radius during slow roll, an important assessment in the analysis below.

 In terms of these definitions and eqn. (\ref{equ}), we have that during the fast roll stage
\be \dot{\Phi}(t) =\frac{\dot{\Phi}_{sr} }{ {a}^3(t)}   \,. \label{fit}\ee Introducing
\be H^2_{sr} \equiv \frac{V_{sr} }{3M^2_{Pl}}\,, \label{Hsr}\ee Friedmann's equation becomes
\be \frac{\dot{ {a}}(t)}{a(t)} = H_{sr} \Bigg[1+\frac{\epsilon_V}{3\, {a}^6(t)}\Bigg]^{1/2} \,. \label{frie}\ee This equation for the scale factor can be readily integrated to yield the solution
\be  {a}(t) = \Bigg[\Bigg( \frac{\epsilon_V}{3}\Bigg)^{1/2} \sinh[\theta(t)] \Bigg]^{1/3}~~;~~\theta(t)=\theta_0+3H_{sr}t \label{tila}\ee where $\theta_0$ is an integration constant chosen to be
\be e^{-\theta_0} = \sqrt{\frac{\epsilon_V}{12}} \,,\label{teta0} \ee so that at long time $a(t) = e^{H_{sr}t}$. The slow roll stage begins when $a(t_{sr}) =1$ which corresponds
   to the value of $\theta_{sr}=\theta(t_{sr})$ given by
\be e^{-\theta_{sr}}= f\big(\frac{\epsilon_V}{3}\big) \label{thetasr} \ee where to simplify notation later we defined
\be f(s) = \frac{\sqrt{s}}{1+\sqrt{1+s}}\,. \label{fofs}\ee Introducing the dimensionless ratio of kinetic to potential contributions at the initial time $t_i$
 \be \frac{\dot{\Phi}^2_i}{2V_{sr}} = \kappa \label{kapa} \,, \ee and \emph{assuming} that the potential does not vary very much between the initial time and the onset of  slow roll (this is quantified further in ref.\cite{lellor}), it follows from (\ref{equ})  that
\be   a^6_i = \frac{\dot{\Phi}^2_{sr}}{2V_{sr}\kappa} = \frac{\epsilon_V}{3\kappa} \label{equa2}\ee where we have used (\ref{fidotsr}). Combining this result with (\ref{tila}), we find that at the initial time $\theta_i=\theta(t_i)$ is given by
\be e^{-\theta_i} = f(\kappa)\,.\label{tetai}\ee Let us introduce
\be \varepsilon(t) = - \frac{\dot{H}}{H^2} = \frac{\dot{\Phi}^2}{2M^2_{Pl}H^2} = \frac{\epsilon_V}{a^6(t)+\frac{\epsilon_V}{3}} \label{epsilon}\ee where we have used the results (\ref{solinfla},\ref{fidotsr},\ref{equ})   from which it is clear that for $\epsilon_V \ll 1$ the slow roll stage begins at $a=1$ when $\varepsilon = \epsilon_V + \mathcal{O}(\epsilon^2_V)$. With $a(t)$ given by (\ref{tila}), it follows that
\be \varepsilon(t) = \frac{3}{\cosh^2[\theta(t)]}\,, \label{vare2} \ee therefore   $0 \leq \varepsilon \leq 3$,    and
\be H(t) = \frac{H_{sr}}{\tanh[\theta(t)]}\,. \label{hubble}\ee

The acceleration equation written in terms of $\varepsilon(t)$ is given by
\be \frac{\ddot{a}}{a} = H^2(t)(1-\varepsilon(t))\,, \label{acce}\ee so that the inflationary stage begins when $\varepsilon(t) =1$. At the initial time
\be \varepsilon(t_i) = \frac{3\kappa}{1+\kappa} \label{vareini} \ee hence, for $\kappa > 1/2$ the early stage of expansion is deccelerated and inflation begins when $\varepsilon(t_{inf}) =1$.

 It   proves convenient to introduce the variable
 \be x(t) = e^{-\theta(t)/3} = \Big[\frac{\epsilon_V}{12}\Big]^{1/6}\,e^{-H_{sr}t}  \,, \label{xoft} \ee  with
\be x_i\equiv x(t_i) = [f(\kappa)]^{1/3}~~;~~x_{sr}\equiv x(t_{sr}) = [f(\epsilon_V/3)]^{1/3}\,. \label{xs}\ee where $f(s)$ is given by eqn. (\ref{fofs}),  and write $a, H,\varepsilon$ in terms of this variable leading to
\be a(x) = \Big[\frac{\epsilon_V}{12} \Big]^{1/6}\,\frac{\big[1-x^6\big]^{1/3}}{x} \,,\label{aofx}\ee
\be H(x)= H_{sr} \frac{\big[1+x^6\big]}{\big[1-x^6\big]} \label{Hofx}\,,\ee
\be \varepsilon(x) = \frac{12\,x^6}{\big[1+x^6\big]^2} \label{epsiofx}\,.\ee

 Conformal time $\eta(t)$ defined to vanish as $t\rightarrow \infty$  is given by
\bea \eta(t)  & = &  \int^{t}_{\infty} \frac{dt'}{a(t')} = \int^{a(t)}_{\infty} \frac{da}{a^2\,H(a)} \nonumber \\ & = & -\frac{1}{a(t)H(t)} +\int^{t}_{\infty} \varepsilon(t')\frac{dt'}{a(t')} \label{etadef}\eea where we integrated by parts and used the definition of $\varepsilon$ given by eqn. (\ref{epsilon}). Adding and subtracting $\epsilon_V$  we find

\be \eta(t)  = -\frac{1}{a(t)H(t)(1-\epsilon_V)}+ \frac{\epsilon_V}{ \big(1-\epsilon_V\big)}\int_{\infty
}^{t}\,\Big[\frac{\varepsilon(t')}{\varepsilon_V}-1\Big] \frac{dt'}{a(t')}    \,, \label{etadef2} \ee

The argument of the integrand in the second term in (\ref{etadef2}) vanishes to leading order in $\epsilon_V,\eta_V$  in the slow roll phase (when $t>t_{sr}$). Therefore, during slow roll, $\eta = -1/aH(1-\epsilon_V)$. For numerical purposes it is convenient to write $\eta$ in terms of the variable $x$ (\ref{xoft}), it is given by
\bea \eta (x) & = &  -  \frac{1}{H_{sr}\,\big(1-\epsilon_V\big)} \, \Big( \frac{12}{\epsilon_V}\Big)^{1/6} \Bigg\{ \frac{x(1-x^6)^{2/3}}{(1+x^6)} \nonumber \\ &+ & \epsilon_V \,\int^{x}_{x_{sr}} \frac{dy}{[1-y^6]^{1/3}}\,
\Bigg[ \frac{12}{\epsilon_V} \frac{y^6}{(1+y^6)^2}-1\Bigg] \Bigg\}\,. \label{etaofx}\eea

The number of e-folds between the initial time $t_i$ and a given   time $t$ is given by
\be N_e (t;t_i) = \int_{t_i}^{t} H(t')\, dt' = \frac{1}{3}\,\ln\Bigg[    \sqrt{\kappa} \, \frac{(1-x^6(t))}{2x^3(t)}\Bigg]\,,  \label{Ne} \ee with a total number of e-folds between the beginning of the fast roll stage at $t=t_i$ and the onset of slow roll at $t_{sr}$ given by
 \be N_{e}(t_i;t_{sr})=  \frac{1}{6}\,\ln\Big[   \frac{3 {\kappa}}{\epsilon_V} \Big]\,.  \label{Netot} \ee

\begin{figure}[h!]
\begin{center}
\includegraphics[height=3.2in,width=3.0in,keepaspectratio=true]{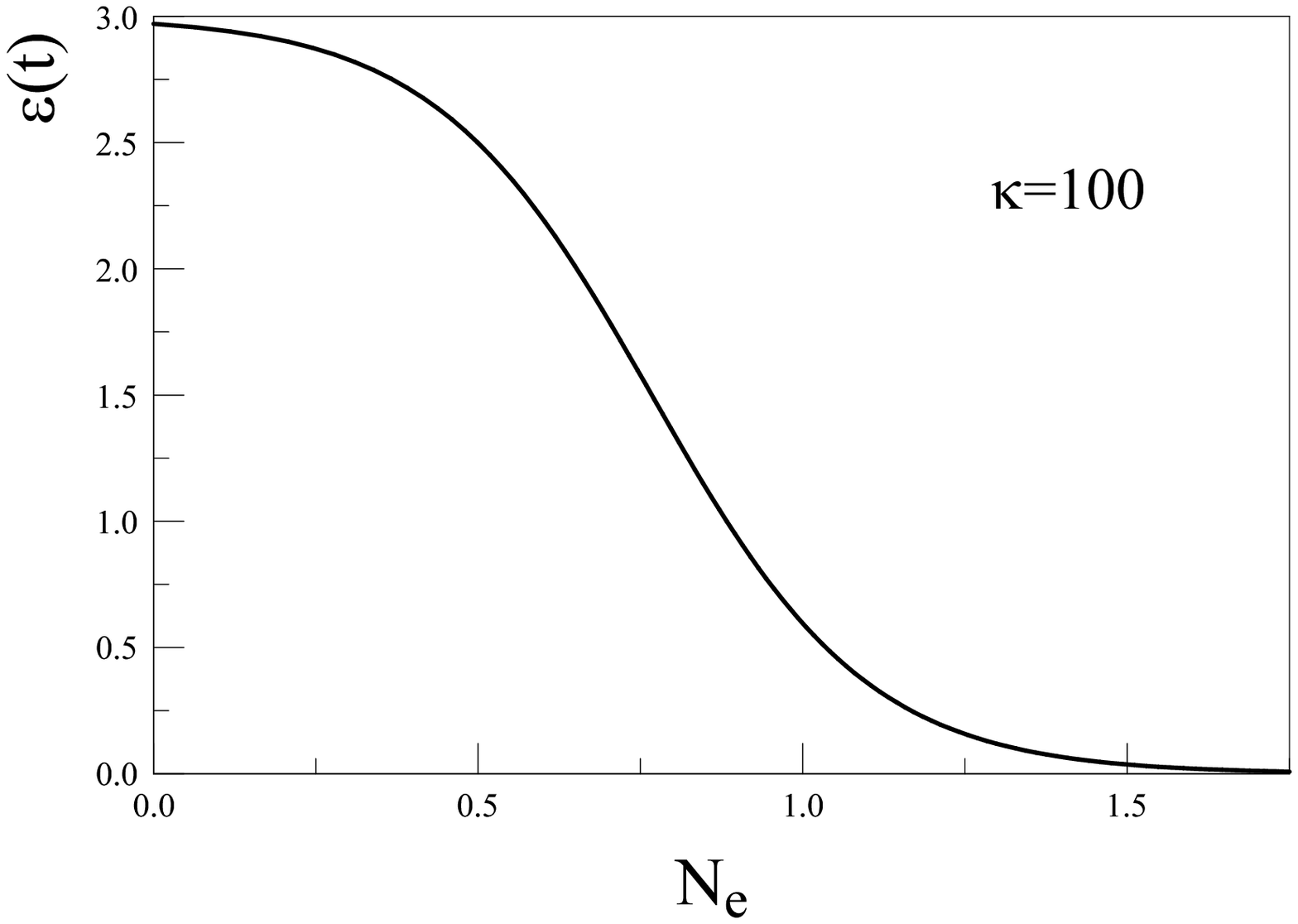}
\includegraphics[height=3.2in,width=3.0in,keepaspectratio=true]{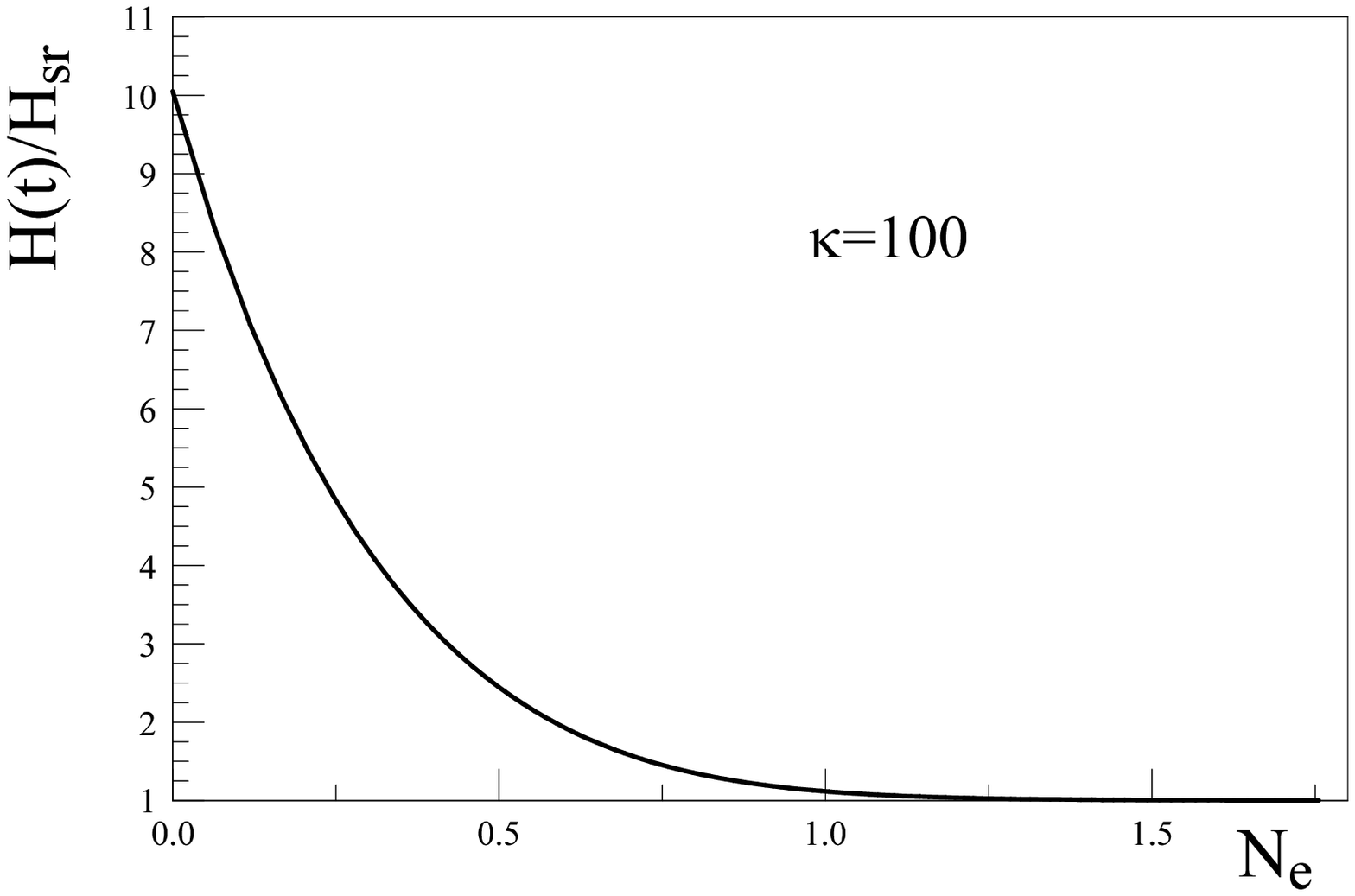}
\caption{$\varepsilon(t)$   and $H(t)/H_{sr}$   as a function of the number of e-folds from the beginning of fast roll, for $\kappa=100$ for $\epsilon_V=0.008$. Inflation starts at $N_e \simeq 0.5$, slow roll starts at $N_e \lesssim 1.75$. }
\label{eNe}
\end{center}
\end{figure}

 Fig. (\ref{eNe}) shows $\varepsilon$ as a function of $N_e$ for $\kappa = 100,\epsilon_V = 0.008$, inflation begins at $N_e \simeq 0.5-0.8$ and slow roll begins at $N_e\simeq 1.37-1.75$. We find that this is the typical behavior for $1 \leq \kappa \leq 100$,  namely for a wide range of fast roll initial conditions, the inflationary stage begins fairly soon $N_{e,inf} \lesssim 1$ and the fast roll stage lasts  $\lesssim 1.7$ e-folds.

The results above are the leading order contributions in $\epsilon_V$ during the fast roll stage, higher orders are studied systematically in ref.\cite{lellor}.

The latest results from the Planck collaboration\cite{planck} confirm a $5-10\%$ suppression of power for $l\lesssim 40$ with $2.5-3\,\sigma$ significance. Recently in ref.\cite{lellor} a detailed study  of the impact of the fast-roll stage on non-Bunch Davies initial conditions for curvature perturbations and on the suppression of the low multipoles has been reported. The results of this reference are independent of the inflaton potential and suggest that a $5-10\%$ suppression of the quadrupole is consistent with a fast roll stage with a ratio of kinetic to potential contributions $10\lesssim \kappa \lesssim 100$. These results confirm more generally previous results based on particular realizations of the inflaton potential\cite{lindefast,contaldi,boyan3,hectordestri,reviunos}.

\section{Initial conditions from a pre-slow roll stage:}

Our goal is to understand how infrared aspects of light scalar fields with mass $M\ll H$, are modified by the ``fast-roll'' stage, therefore in this and following sections we focus on ``test'' scalar fields, not necessarily the inflaton field.

The quantization of a generic minimally coupled massive scalar field is achieved by writing
\be \phi(\vec{x},\eta) = \frac{1}{C(\eta)}\frac{1}{\sqrt{V}} \sum_{\vec{k}} \left[\alpha_{\vk} \; S (k,\eta)\,e^{i\vec{k}\cdot\vec{x}}+
\alpha^{\dagger}_{\vk} \; S^*(k,\eta)\,e^{-i\vec{k}\cdot\vec{x}}\right] \; ,
\label{expalfa} \ee where the operators $ \alpha_{\vk}, \; \alpha^{\dagger}_{\vk} $   obey the usual canonical commutation relations, and the mode functions $ S_\phi(k,\eta) $ are
solutions of \be \left[\frac{d^2}{d\eta^2}+k^2 -W(\eta)  \right]S(k,\eta) = 0  ~~;~~ W(\eta)  = \frac{C''(\eta)}{C(\eta)}-
M^2 \; C^2(\eta)  \label{phieqn}
\; . \ee

 This is  a
Schr\"odinger equation, with $ \eta $ the coordinate, $ k^2 $ the
energy and $ W(\eta) $ a potential that depends on the coordinate $
\eta $.  The full dynamics of the inflaton field during the fast roll stage yields the potential
 \be W (\eta) = \frac{C''}{C}-M^2 \, C^2(\eta) = a [\ddot{a}+H\dot{a}]-M^2\,a^2(t)  =     2 a^2 H^2 \Big[1-\frac{3}{2}\, \Delta- \frac{\varepsilon(t)}{2}\Big]\,,  \label{poten} \ee where we have introduced
\be \Delta = \frac{M^2}{3H^2} \ll 1 \label{deltadef1}\,. \ee
 During slow roll inflation the potential  $\varepsilon = \epsilon_V$ and
\be a^2(t) H^2(t) = \frac{1}{\eta^2} \,(1+2\epsilon_V)\label{a2H2sr}\ee therefore, during slow roll
 $W(\eta)$ becomes
\be\label{defV}
W(\eta)=
\frac{\nu^2-\frac14}{\eta^2} \; ,
\ee where to leading order in slow roll parameters and $\Delta$
\be \nu  = \frac{3}{2}+
 \epsilon_V -\Delta \,.\label{defnu}
\ee  Therefore during the full dynamics of the inflation including the fast roll stage we write
\be W(\eta)\equiv \mathcal{V}(\eta) + \frac{\nu^2-\frac14}{\eta^2}\label{Wfull}\ee where the potential
\be  \mathcal{V}(\eta) = W(\eta)-\frac{2}{\eta^2}\Big[1+\frac{3\epsilon_V}{2}- \frac{3\Delta}{2}\Big]\,. \label{vpot}\ee

\begin{figure}[h!]
\includegraphics[height=3.2in,width=3.2in,keepaspectratio=true]{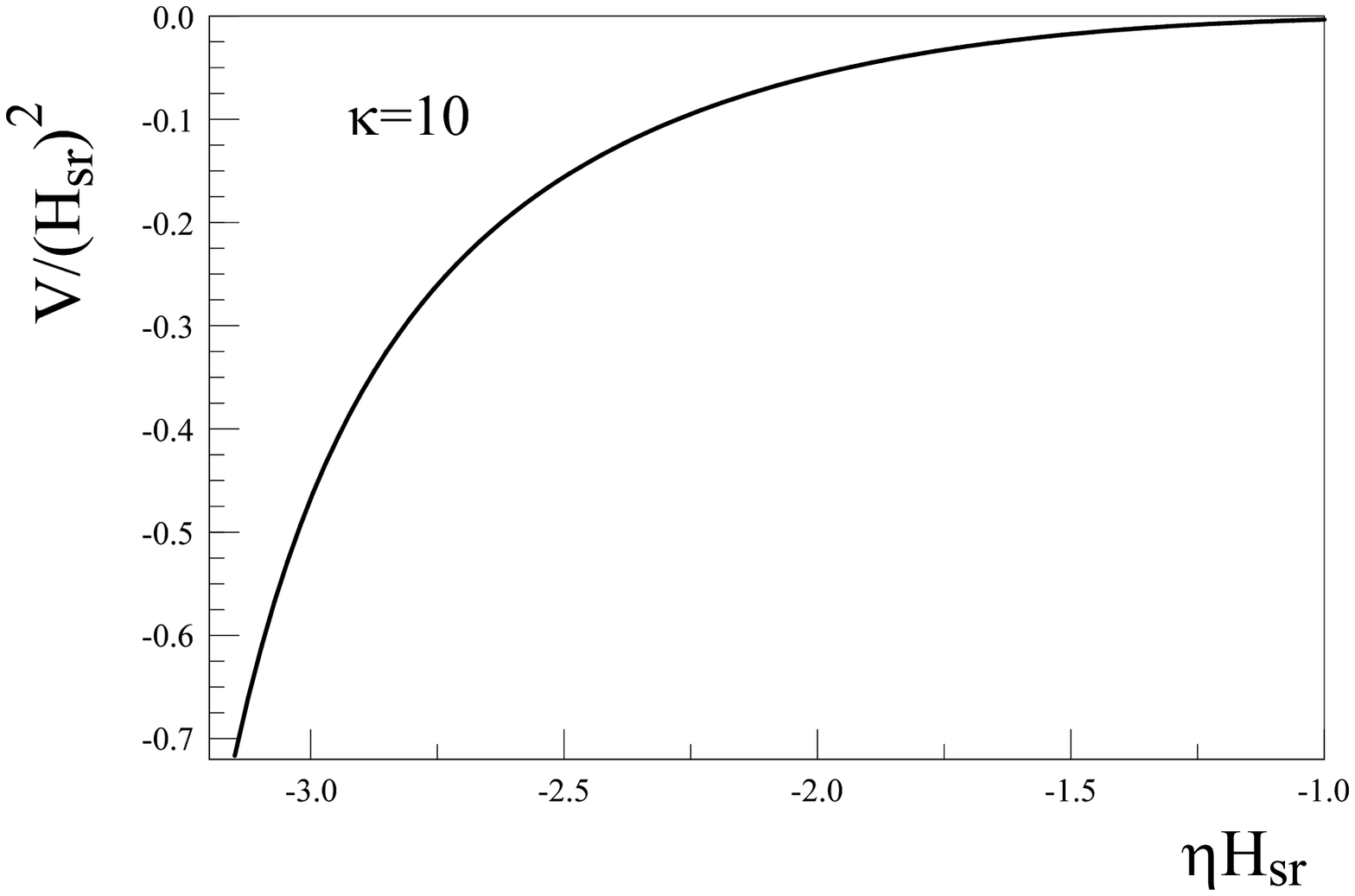}
\includegraphics[height=3.2in,width=3.2in,keepaspectratio=true]{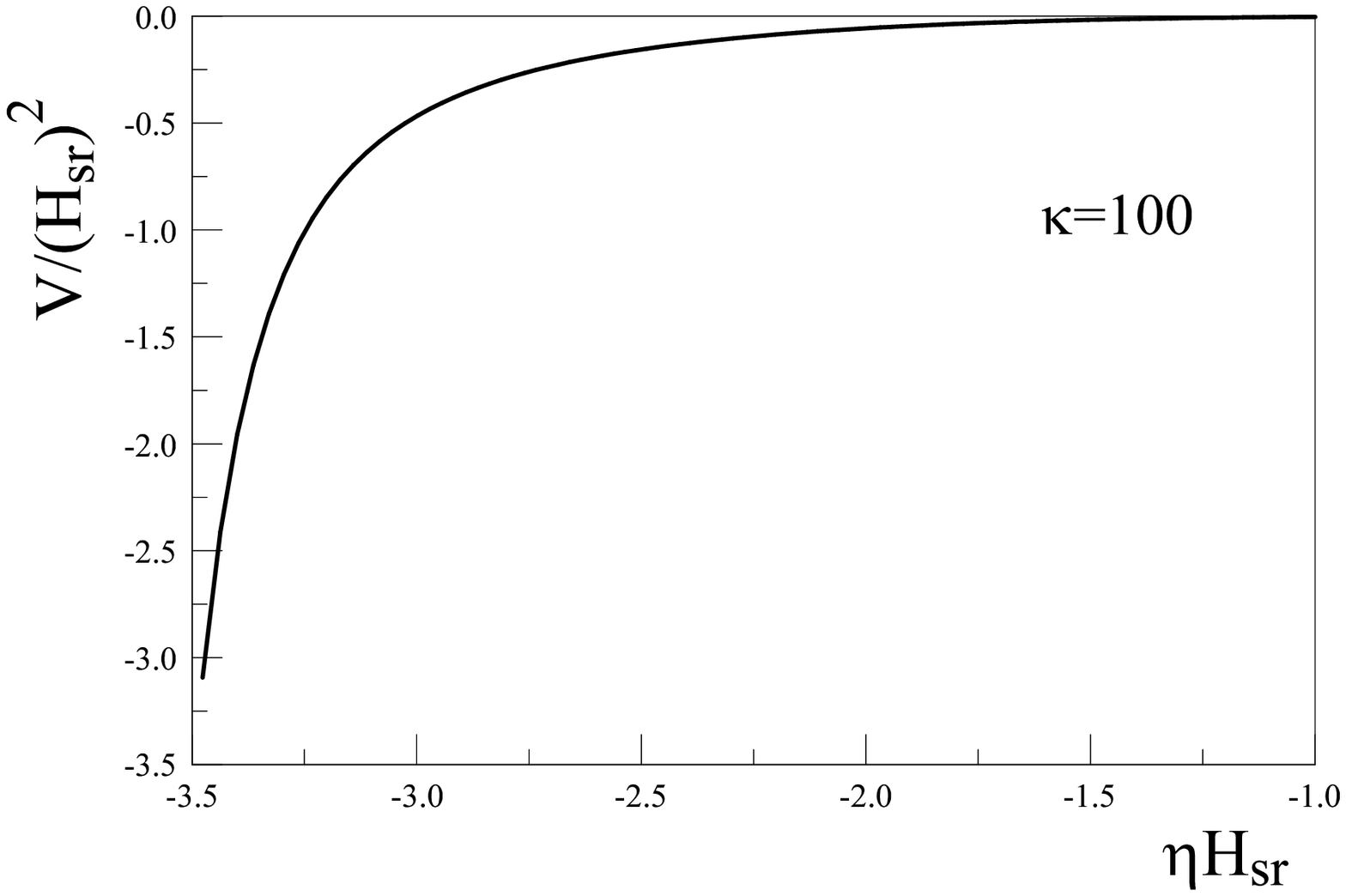}
\caption{Potentials      as a function of $\eta$ from the beginning of fast roll, for $\kappa=10;100~;~\epsilon_V=0.008~;~\Delta=0.01$.  }
\label{fig:potentials}
\end{figure}

The potential is calculated parametrically in terms of the variable $x$ introduced in (\ref{xoft}) and $a,H,\eta$ all functions of $x$ given by the expressions (\ref{aofx},\ref{Hofx},\ref{etaofx}). Figure (\ref{fig:potentials}) shows the typical potentials for $\kappa = 10,100; \epsilon_{V}=0.008;\Delta = 0.01$. We studied the potentials for a wide range of values of $\epsilon_V,\Delta$ and $\kappa$ with qualitatively the same features.

The potentials are always \emph{negative} and qualitatively of the same form with very small variations for fixed $\kappa$   the (negative) amplitude of the potential increases with increasing $\kappa$. For both $\epsilon_V;\Delta \ll 1$ the potential is quite insensitive to their values and is mainly determined by the ratio $\kappa$.

These results are in general agreement with those of refs. \cite{boyan3,hectordestri,reviunos,lasenby} and more recently in ref.\cite{lellor}  a  more detailed analysis confirmed the robustness of the main features of the pre-slow roll stage quite independently of the inflationary potential (provided the potential is smooth enough to be consistent with slow roll).

The solution of the mode equations with Bunch-Davies initial conditions for sub horizon modes obey the condition
\be S(k;\eta) \rightarrow \frac{e^{-ik\eta}}{\sqrt{2k}} ~~;~~ -k\eta \rightarrow \infty \,,\label{BDbc}\ee and up to an overall phase are given  by
\be S(k;\eta)\equiv g_{\nu}(k,\eta) =   \sqrt{\frac{-\pi \eta}{4}} \, H^{(1)}_{\nu} (-k \eta)\; , \label{BDmodes} \ee these mode functions satisfy the Wronskian condition
\be \mathcal{W}[g,g^*] = g^{'}_{\nu}(k,\eta)\,g^*_{\nu}(k,\eta)- g^{*\,'}_{\nu}(k,\eta)\,g_{\nu}(k,\eta) = -i\,. \label{wronskian}\ee

When field quantization is carried out with these mode functions the  vacuum state $|0\rangle_{BD}$ annihilated by the operators $\alpha_{\vec{k}}$ is the Bunch-Davies vacuum. However, the most general solution in the slow-roll regime can be written as
\be S(k;\eta) =  A_k g_{\nu}(k,\eta) + B_k g^*_{\nu}(k,\eta) \,\label{genS}\ee where $A_k;B_k$ are Bogoliubov coefficients. For the creation and annihilation operators to obey standard commutation relations it follows that these general combinations must obey the Wronskian condition
\be \mathcal{W}[S,S^*] = -i = \mathcal{W}[g,g^*] \Big[|A_k|^2-|B_k|^2 \Big] \label{Ws}\ee from which it follows that the Bogoliubov coefficients   must obey the constraint
\be |A_k|^2-|B_k|^2 =1 \,. \label{constraint}\ee

The relation between quantization with the mode functions $S(k;\eta)$ with general initial conditions, and the more familiar Bunch-Davies case with the mode functions $g_\nu$ (\ref{BDmodes})  is obtained from the expansion of the Fourier components of the relevant fields, namely the field can be expanded in either set with corresponding annihilation and creation operators, for example for a scalar field
\be \frac{1}{\sqrt{V}} \sum_{k}  a_{\vk} g_{\nu}(k, \eta) e^{i\vec{k} \cdot \vec{x}} + a^{\dagger}_{\vk} g^{*}_{\nu}(k, \eta) e^{-i\vec{k} \cdot \vec{x}} = \frac{1}{\sqrt{V}} \sum_{k}  \alpha_{\vk} S(k, \eta) e^{i\vec{k} \cdot \vec{x}} + \alpha^{\dagger}_{\vk} S^*(k, \eta) e^{-i\vec{k} \cdot \vec{x}}
\ee where $a_{\vk} |0 \rangle_{BD} = 0$ defines the Bunch-Davies vacuum and $\alpha_{\vk}|0\rangle_\alpha$ defines the vacuum with the general initial conditions. The relation between the creation and annihilation operators is obtained from the Wronskian conditions, it is given by

\be
\alpha_{\vk} = A^*_k a_{\vk} - B^*_k a^{\dagger}_{-{\vk}} ~~;~~ \alpha^{\dagger}_{\vk} = A_k a^{\dagger}_k - B_k a_{-{\vk}}\,. \label{relas}
\ee

The Bogoliubov coefficients have been discussed in the literature\cite{mukh,kolb,riotto2,baumann,giov}   and an interpretation can be furnished by considering the action of the $\alpha$ number operator on the Bunch-Davies vacuum. It is easily shown that

\be
{}_{BD}\langle 0  | \alpha^{\dagger}_k \alpha_k | 0  \rangle_{BD} = |B_k|^2
\ee which suggests the interpretation that $|B_k|$ is the number of $\alpha$-vacuum particles in the Bunch Davies vacuum.

The power spectra for scalar field fluctuations ($\phi$) is given by,
\be \mathcal{P}(k) = \frac{k^3}{2\pi^2} \Bigg|\frac{S(k;\eta)}{C(\eta)}\Bigg|^2 \label{powerspectra}\ee  Evaluating these power spectra a few e-folds after horizon crossing $-k\eta \ll 1$ and using that in this regime $H^{(1)}_{\nu} (-k \eta) \simeq i\,Y_\nu(-k\eta)$ it follows that  for $-k\eta \ll 1$  the general solution of the form (\ref{genS}) is given by
\be S(k;\eta)  =   i\,\sqrt{\frac{-\pi \eta}{4}} \, Y_\nu(-k\eta)\big[A_k-B_k \big]\; , \label{BDmodeslargeeta} \ee therefore the power spectra becomes

\be \mathcal{P}(k) =   \mathcal{P}^{BD}(k)\,\mathcal{T}(k)  \,, \label{transfer} \ee where $\mathcal{P}^{BD}(k)$ are the power spectra for Bunch-Davies modes $g_\nu(k;\eta)$, namely for  $A_k =1;B_k=0$, and
\be \mathcal{T}(k) = \big|A_k - B_k\big|^2 \label{tfun}\ee is a transfer function that encodes the non-Bunch-Davies initial conditions for the respective perturbations.

The main question is precisely what is the origin of $\mathcal{T}(k)$ and what are the properties for small and large $k$.

 In   references \cite{boyan3,hectordestri,reviunos,lasenby} and more recently in ref.\cite{lellor} the modifications of the mode equations during the fast-roll stage,  where the  stage just prior to   slow roll   was kinetic dominated, were invoked as a \emph{possible} origin of the Bogoliubov coefficients $A_k~,~B_k$.

Here we pursue this line of argument and consider this possibility in detail, in particular focusing on the superhorizon limit of the transfer function $\mathcal{T}(k)$ (\ref{tfun}) for light ``test'' scalar fields, namely with $\Delta \ll 1$.

 The full dynamical evolution of the inflaton leads to a modification of the mode equations (\ref{phieqn}) where $W(\eta)$ is now given by (\ref{Wfull}) in terms of the potential $\mathcal{V}(\eta)$. As shown in figure (\ref{fig:potentials}) this potential is localized in $\eta$ in a narrow range prior to the slow roll phase\cite{boyan3,hectordestri,reviunos,lasenby,lellor}, namely in the mode equations (\ref{phieqn}) $W(\eta)$ is written as
\be W(\eta) = \mathcal{V}(\eta) + \frac{\nu^2-1/4}{\eta^2} ~~;~~ \mathcal{V}(\eta) = \Bigg\{\begin{array}{l}
                \neq 0 ~\mathrm{for} ~-\infty < \eta < {  \eta_{sr}} \\
                0 ~\mathrm{for} ~ {  \eta_{sr}} < \eta   \\
              \end{array} \,.\label{Vpot} \ee where $\eta_{sr}$ determines the beginning of the slow roll stage when $\epsilon_V~,~\eta_V \ll 1$ (see figure (\ref{fig:potentials})).

Rather than studying the behavior of the Bogoliubov coefficients numerically for different values of the parameters, we now exploit the similarity with a quantum mechanical potential problem and implement methods from the quantum theory of scattering to obtain the \emph{general} behavior on $\mathcal{T}(k)$  for small and large wavevectors  based solely on the fact that the potential is \emph{negative} and localized. These are \emph{generic} features of the potentials $\mathcal{V}(\eta)$ as consequence of the brief fast roll stage prior to slow roll.

The mode equation (\ref{phieqn}) can now be written as
\be  \Big[\frac{d^2}{d\eta^2}+k^2-\frac{\nu^2-1/4}{\eta^2}\Big]S(k;\eta) = \mathcal{V}(\eta) S(k;\eta) \,, \label{inteq} \ee which can be converted into an integral equation via the retarded Green's function $G_k(\eta,\eta')$ obeying
\be \label{green}
\left[\frac{d^2}{d\eta^2}+k^2-\frac{\nu^2-\frac14}{\eta^2}
 \right]G_k(\eta,\eta') = \delta(\eta-\eta') ~~;~~ G_k(\eta,\eta') =0~
\mathrm{for}~\eta'>\eta \,.
\ee
This Green's function  is given by
\be \label{Gret}
G_k(\eta,\eta') = i \left[g_\nu(k;\eta) \; g^*_\nu(k;\eta')-
g_\nu(k;\eta') \; g^*_\nu(k;\eta) \right] \Theta(\eta-\eta') \quad ,
\ee
where $ g_\nu(k;\eta) $ is given by eq.(\ref{BDbc}). The solution of (\ref{inteq}) with boundary conditions corresponding to Bunch-Davies modes deep inside the horizon obeys the Lippman-Schwinger integral equation familiar from scattering theory,
\be \label{sol}
S(k;\eta)= g_\nu(k;\eta) +
\int^{0}_{-\infty} G_k(\eta,\eta') \; \mathcal{V}(\eta') \;
S(k;\eta') \; d\eta' \;.
\ee With the Green's function given by (\ref{green}) this solution can be written as
\be S(k;\eta) = A_k(\eta) \,g_\nu(k;\eta) + B_k(\eta)\,g^*_\nu(k;\eta) \,, \label{soluLip}\ee where
\bea  A_k(\eta) & = & 1+  i \int^{\eta}_{-\infty}
   \mathcal{V}(\eta')\,g^*_\nu(k;\eta') \,S(k;\eta')  \, d\eta' \label{aofketa} \\
 B_k(\eta) & = & -  i \int^{\eta}_{-\infty}
   \mathcal{V}(\eta')\,g_\nu(k;\eta') \,S(k;\eta')  \, d\eta' \,.\label{bofketa}\eea For a potential $\mathcal{V}(\eta)$ that is localized prior to the slow roll stage (see fig. \ref{fig:potentials}), and for $\eta > \eta_{sr}$ we can safely replace the upper limit of the integrals $\eta \rightarrow 0$  and during the slow roll stage the solution (\ref{soluLip}) becomes
   \be S(k;\eta) = A_k \,g_\nu(k;\eta) + B_k\,g^*_\nu(k;\eta)~~;~~A_k \equiv A_k(\eta =0)~;~B_k \equiv B_k(\eta =0)\,. \label{Ssr}\ee This expression clearly suggests that mode functions with general initial conditions follow from pre-slow-roll stage wherein the inflaton zero mode undergoes rapid dynamical evolution. Refs.\cite{lellor} provides a thorough numerical study of the potential independently of the inflaton potential (see figs. in this reference).

    We now pursue an analytic understanding of the transfer function $\mathcal{T}(k)$ both for super and subhorizon modes quite generically without specifying particular values of $\kappa;\epsilon_V;\Delta$ but based \emph{solely} on the fact that the potential $\mathcal{V}(\eta)$ is localized and negative.

       We first note that the $\eta$ dependent Bogoliubov coefficients (\ref{aofketa},\ref{bofketa}) satisfy the
   relation
   \be g_\nu (k;\eta) A'_{k}(\eta)+ g^*_\nu (k;\eta) B'_{k}(\eta) =0 \,,\label{rel1} \ee which implies
   the following relation  between Wronskians
   \be \mathcal{W}[S,S^*] = \mathcal{W}[g_\nu,g^*_\nu]\,\Big(|A_k(\eta)|^2-|B_k(\eta)|^2\Big)\,. \label{rel2}\ee valid at \emph{all times} not only during slow roll.

   Secondly, inserting the relation (\ref{soluLip}) into the   equations (\ref{aofketa},\ref{bofketa}) leads to the coupled Fredholm integral equations
   \bea     A_k(\eta) & = & 1+  i \int^{\eta}_{-\infty} \Big\{ C(k;\eta') A_k(\eta') + D(k;\eta') B_k(\eta')\Big\}    \, d\eta' \label{freda} \\
 B_k(\eta) & = & -  i \int^{\eta}_{-\infty}\Big\{ C(k;\eta')B_k(\eta') + D^*(k;\eta') A_k(\eta')\Big\}  \, d\eta' \,,\label{fredb}\eea where the coefficient functions are given by
   \be C(k;\eta) =
     |g_\nu(k;\eta)|^2 \,\mathcal{V}(\eta) ~~;~~D(k;\eta) = \big(  g^*_\nu(k;\eta)  \big)^2 \,\mathcal{V}(\eta)\,. \label{albet}\ee Upon taking derivatives with respect to conformal time we find the coupled differential equations
     \bea  A'_k(\eta) & = &   i  C(k;\eta) A_k(\eta') + i D(k;\eta) B_k(\eta) ~~;~~ A_k(k;-\infty) = 1 \label{aprima}\\
    B'_k(\eta) & = & -  i  C(k;\eta)B_k(\eta) -i D^*(k;\eta) A_k(\eta) ~~;~~ B_k(k;-\infty) =0 \,. \label{bprima}\eea It is straightforward to confirm that these equations lead to the result
    \be \frac{d}{d\eta} \Big( |A_k(\eta)|^2 -|B_k(\eta)|^2 \Big)=0 \,,\label{constAB}\ee which combined with the initial conditions in eqns. (\ref{aprima},\ref{bprima}) yield the $\eta$-independent result
    \be |A_k(\eta)|^2 -|B_k(\eta)|^2 =1 \,.\label{WAB1} \ee Along with the relation (\ref{rel2}) this result implies that  $\mathcal{W}[S,S^*]=-i$, namely the fields quantized with the Bunch-Davies modes and the modes $S(k;\eta)$ which are determined by the pre-slow roll stage are related by a canonical transformation.

    Writing the coefficients $C(k;\eta);D(k;\eta)$ explicitly in terms of Bessel functions, it follows that
    \be C(k;\eta) + D^*(k;\eta) = \Big( \frac{-\pi \eta}{2} \Big)\,\mathcal{V}(\eta) \Big[ J^2_\nu(-k\eta)+i  J_\nu(-k\eta)Y_\nu(-k\eta) \Big]\,. \label{CplusD}\ee
    \be C(k;\eta) - D^*(k;\eta) = \Big( \frac{-\pi \eta}{2} \Big)\,\mathcal{V}(\eta) \Big[ Y^2_\nu(-k\eta)-i  J_\nu(-k\eta)Y_\nu(-k\eta) \Big]\,. \label{CminD}\ee

The coupled set of linear differential equations (\ref{aprima},\ref{bprima}) is difficult to solve analytically in general although the system is amenable to a straightforward numerical integration. However analytical progress can be made in two limits: a) the superhorizon limit  $-k\eta \rightarrow 0$, b) subhorizon modes $-k\eta \gg 1$.

\vspace{2mm}

\textbf{Superhorizon modes:} For modes that crossed the horizon prior to the onset of the slow-roll phase and either during or prior to the stage where the inflaton field is evolving rapidly

    \be J_\nu(-k\eta) \simeq \frac{\big(-k\eta/2 \big)^\nu}{\nu\,\Gamma(\nu)} ~~;~~ Y_\nu(-k\eta) \simeq -\frac{\Gamma(\nu)}{\pi}\, {\big(-k\eta/2 \big)^{-\nu}} \,. \label{suphorbes}\ee It proves convenient to define the combinations
    \be F_{ \pm}(k;\eta) = A_k(\eta)\pm B_k(\eta)\,, \label{Fpm}\ee obeying the coupled equations
    \bea && F'_-(k;\eta)-\gamma(\eta) F_-(k;\eta) = i\pi \nu \gamma(\eta) J^2_\nu(-k\eta) F_+(k;\eta) \label{Fmin}\\
     && F'_+(k;\eta)+\gamma(\eta) F_+(k;\eta) = i\pi \nu \gamma(\eta) Y^2_\nu(-k\eta) F_-(k;\eta)\,, \label{Fplus} \eea where we have introduced
     \be \gamma(\eta) = \Big( \frac{-\eta}{2\nu}\Big) \, \mathcal{V}(\eta)\,. \label{gamma} \ee The equations (\ref{Fmin},\ref{Fplus}) can be simplified by writing
     \be F_{ \pm}(k;\eta) = h_{\pm}(\eta) \,f_{\pm}(k;\eta)~~;~~ h_{\pm}(\eta) = \exp\Big\{\mp \int^{\eta}_{-\infty} d\eta' \gamma(\eta')\Big\}\,,  \label{aches} \ee and defining
     \be \tilde{j}(k;\eta)  \equiv \pi \nu J^2_{\nu}(-k\eta) \,h^2_{+}(\eta) ~~;~~ \pi \nu Y^2_{\nu}(-k\eta) \,h^2_{-}(\eta) = \frac{1}{\tilde{j}(k;\eta)}\,, \label{jtil}\ee where we have used the limiting form (\ref{suphorbes}) for superhorizon modes. With these definitions one finds the following set of coupled equations for the real and imaginary parts,
\bea \mathrm{Re}\,f'_-(k;\eta)  & = &  - \gamma(\eta)\,\tilde{j}(k;\eta)~ \mathrm{Im}\,f_+(k;\eta) \label{Refmin} \\\mathrm{Re}\,f'_+(k;\eta)  & = &  - \frac{\gamma(\eta)}{\tilde{j}(k;\eta)}\, \mathrm{Im}\,f_-(k;\eta) \,, \label{Refplus} \eea

\bea \mathrm{Im}\,f'_-(k;\eta)  & = &   \gamma(\eta)\,\tilde{j}(k;\eta)~ \mathrm{Re}\,f_+(k;\eta) \label{Imfmin} \\\mathrm{Im}\,f'_+(k;\eta)  & = &   \frac{\gamma(\eta)}{\tilde{j}(k;\eta)}\, \mathrm{Re}\,f_-(k;\eta) \,, \label{Imfplus} \eea with the initial conditions
\be \mathrm{Re}f_{\pm} (k;\eta \rightarrow -\infty) \rightarrow 1 ~~;~~  \mathrm{Im}f_{\pm} (k;\eta \rightarrow -\infty) \rightarrow 0\,. \label{icfs}\ee
Given the potential $\mathcal{V}(\eta)$ this set of equations lends itself to a simple numerical integration. However we can pursue further analytical understanding by writing them into an equivalent set of integral equations as follows: formally integrating (\ref{Refplus},\ref{Imfplus}) with the initial condition (\ref{icfs}) and introducing the result into the equations for (\ref{Refmin},\ref{Imfmin}), we integrate  with the initial condition (\ref{icfs})   and obtain

\be \mathrm{Re}f_-(k;\eta) = 1- \int^\eta_{-\infty} d\eta' \gamma(\eta') \tilde{j}(k;\eta') \int^{\eta'}_{-\infty} d\eta'' \frac{\gamma(\eta'')}{\tilde{j}(k;\eta'')}\,\mathrm{Re}f_-(k;\eta'') d\eta''
\label{inteqRefmin}\ee

\be \mathrm{Im}f_-(k;\eta)  = \int^\eta_{-\infty} d\eta' \gamma(\eta') \tilde{j}(k;\eta')  - \int^\eta_{-\infty} d\eta' \gamma(\eta') \tilde{j}(k;\eta') \int^{\eta'}_{-\infty} d\eta'' \frac{\gamma(\eta'')}{\tilde{j}(k;\eta'')}\,\mathrm{Im}f_-(k;\eta'') d\eta''\,. \label{inteqImfmin}\ee

Inserting the solutions to these integral equations into equations (\ref{Refmin},\ref{Imfmin}) yield the solutions for $f_+(k;\eta)$.

We can glean several important features from the integral equations (\ref{inteqRefmin},\ref{inteqImfmin}):
\begin{itemize}
\item{  $\mathrm{Re}f_-(k;\eta ) $    has a   smooth $k\rightarrow 0 $ limit as the factors $k^{2\nu}$ cancel between the $\tilde{j}$ in the numerator and denominator in the integral equations. Using the small argument expansion of Bessel functions we find that in the long-wavelength limit  \be \mathrm{Re}f_-(k  ;0)  \simeq  \mathrm{Re}f_-(0 ;0) + \mathcal{O}(k^2) +\cdots \, . \label{smallkRe}\ee where $\mathrm{Re}f_-(0 ;0)$ is finite.

    Since $\tilde{j}(k;\eta) \propto k^{2\nu}$ one notes that rescaling $\mathrm{Im}f_-(k;\eta ) \equiv k^{2\nu} \,\mathrm{Im}\tilde{f}_-(k;\eta )$, it follows from eqn. (\ref{inteqImfmin}) that $\mathrm{Im}\tilde{f}_-(k;\eta )$ has a finite limit as $k\rightarrow 0$ therefore we find that in the long wavelength limit
    \be \mathrm{Im}f_-(k;\eta ) \simeq \mathcal{C}\,k^{2\nu} \Big[1+ \mathcal{O}(k^2) + \cdots \Big] \,. \label{smallkIm}\ee where $\mathcal{C}$ is a finite constant, therefore $\mathrm{Im}[A_{k=0}(\eta)-B_{k=0}(\eta)]=0$.
     From the result $|A_k(\eta)|^2-|B_k(\eta)|^2 =1$ this implies that   the real part $\mathrm{Re}[A_{k=0}-B_{k=0}]$ can \emph{never vanish}. Because of the initial condition this combination begins positive $(=1)$ in the early past and \emph{always remains positive} and the double integral in (\ref{inteqRefmin}) is manifestly \emph{positive} and finite leading to the conclusion that
        \be \mathrm{Re}f_-(0;0) < 1 ~~;~~\mathrm{Im}f_-(0;0)=0 \,. \label{asiReFmin}\ee

      From the result (\ref{smallkIm}) above, and inserting this result in eqn. (\ref{Refplus}) we find that $\mathrm{Re}f_+(k;\eta ) $ features a smooth long-wavelength limit with $\mathrm{Re}f_+(0;0 ) $ a finite constant. Inserting the result  that $\mathrm{Re}f_-(k;\eta ) $ is a regular function approaching a constant in the  long-wavelength limit it follows that  $\mathrm{Im}f_+(k;\eta) \propto k^{-2\nu}$ and features an infrared divergence in the long-wavelength limit. These results for $f_+(k;\eta)$ imply that in the long wavelength limit the \emph{sum}
      \be A_k + B_k \propto i\,k^{-2\nu} \,. \label{sumita}\ee It is important to recognize how, in view of this result, the identity $|A_k(\eta)|^2-|B_k(\eta)|^2 =1$ is fulfilled in the long wavelength limit: from the results $\mathrm{Im}f_-(0;\eta) =0$ and the long wavelengh limit $\mathrm{Im}f_+(k;\eta) \propto k^{-2\nu}$ it follows that in this limit  $[\mathrm{Im}A_k(\eta)]^2 = [\mathrm{Im}B_k(\eta)]^2 \propto k^{-4\nu} $ and $[\mathrm{Re}A_k(\eta)]^2 \simeq \mathcal{O}(1)~;~ [\mathrm{Re}B_k(\eta)]^2 \simeq \mathcal{O}(1)$ from which it follows that $ |A_k(\eta)|^2-|B_k(\eta)|^2 \simeq \mathcal{O}(1)$, namely the singular long wavelength behavior in the imaginary parts of the Bogoliubov coefficients cancel out in the long-wavelength limit, leaving only the regular contributions in this limit.

       During the slow-roll, near de Sitter stage the mode functions become
          \be S(k;\eta) = \frac{1}{2}\sqrt{-\pi\,\eta}\,\Big[ (A_k+B_k)\,J_\nu(-k\eta) + i (A_k-B_k)\,Y_\nu(-k\eta) \Big] \,   \label{modids}\ee in the long-wavelength and long time limit, with the result that $A_k+B_k \propto i k^{-2\nu}$ and $A_k - B_k \simeq \mathcal{O}(1)$, it follows that
          \be S(k;\eta) \simeq \Big[a \, k^{-\nu}   (-\eta)^{\frac{1}{2}+\nu} + b \, k^{-\nu} \, \Big(A_{k=0}-B_{k=0}\Big) (-\eta)^{\frac{1}{2}-\nu}\Big] \label{Sasiketa} \,, \ee where $a,b$ are coefficients of $\mathcal{O}(1)$. Hence, although both terms are of the same order $\propto k^{-\nu}$ in the long wavelength limit, it is the second term that dominates well after horizon crossing and the power spectrum is determined by this term as anticipated above, see the discussion leading up to eqns. (\ref{transfer},\ref{tfun}). In summary for long-wavelength modes at long time $\eta \rightarrow 0$ the mode functions can be approximated as
          \be S(k;\eta) \simeq \frac{-i\,\Gamma(\nu)}{2\pi} (A_k-B_k)\,\sqrt{-\pi\,\eta}~ \Big(\frac{2}{-\eta }\Big)^{\nu}\,k^{-\nu}   \,  . \label{modlwlate}\ee  This result will be used in the analysis of infrared correlations in the next sections. }

           \item{The above results combined with equations  (\ref{Fpm}) and (\ref{aches}) lead to
      \be \mathrm{Re}[A_{k=0}(0)-B_{k=0}(0)] = \exp\Big\{  \int^{0}_{-\infty} d\eta' \gamma(\eta')\Big\} \,\mathrm{Re}f_-(0;0)~~;~~ \mathrm{Im}[A_{k=0}(0)-B_{k=0}(0)]=0 \,, \ee
      hence,
      \be \mathcal{T}(0) = \exp\Big\{2  \int^{0}_{-\infty} d\eta' \gamma(\eta')\Big\} \,\big[\mathrm{Re}f_-(0;0)\big]^2 \label{finiT0}\,.\ee
      Therefore for an \emph{attractive potential} $\mathcal{V}(\eta) < 0$ it follows that $\gamma(\eta) <0$ and
      \be   \mathcal{T}(0) <1 \,, \label{attractiveT0}\, \ee namely, for an \emph{attractive potential}  the long wavelength limit of the initial condition transfer function is smaller than $1$ entailing a \emph{suppression} of the power spectrum at long wavelengths. Since $\big[\mathrm{Re}f_-(0;0)\big]^2 \leq 1$ for the case of attractive potentials as found for a fast-roll stage\cite{boyan3,hectordestri,reviunos,lellor} an \emph{upper bound} for the superhorizon limit of the initial condition transfer function is
       \be \mathcal{T}(0) \leq  \exp\Big\{2  \int^{0}_{-\infty} d\eta' \gamma(\eta')\Big\} \, .\label{upbouT0} \ee

      This analysis confirms more generally the numerical results obtained in refs.\cite{boyan3,hectordestri,reviunos,lellor}. Furthermore using the small argument approximation of the Bessel functions with non-integer $\nu$  the integral equations (\ref{inteqRefmin},\ref{inteqImfmin}) clearly show that
      \be \mathcal{T}(k) \simeq \mathcal{T}(0)+ \mathcal{O}(k^2) +\cdots \label{expaT} \ee namely has a power series expansion in $k$ at long wavelengths. }

\end{itemize}

     \vspace{2mm}

     \textbf{Subhorizon modes:}
     For modes that remain inside the Hubble radius throughout inflation $-k\eta \gg 1$ the integral equation (\ref{sol}) can be consistently solved in a Born series. In the first Born approximation we replace $S(k;\eta) = g_\nu(k;\eta)$ in the integral equation (\ref{sol}) leading to
     \bea  A_k(\eta) & \simeq & 1+  \frac{i}{2k} \int^{\eta}_{-\infty}
   \mathcal{V}(\eta')   \, d\eta' \label{aofketaB} \\
 B_k(\eta) & \simeq & -  \frac{e^{-i\pi \nu}}{2k} \int^{\eta}_{-\infty}\,e^{-2ik\eta'}
   \mathcal{V}(\eta')   \, d\eta' \,.\label{bofketaB}\eea where we have used that for subhorizon modes $g_\nu(k;\eta) \rightarrow e^{-i\frac{\pi}{2}(\nu+1/2)}/\sqrt{2k}$. The subhorizon limit of the coefficient $B_k$ is strongly suppressed because the Fourier transform of the localized potential $\mathcal{V}$ falls off very fast   for large $k$ as a consequence of the Riemann-Lebesgue lemma. An integration by parts dropping the surface terms because a) for large k the integrand at the lower limit averages out to zero and b) for $\eta > \eta_{sr}$  the integrand vanishes at the upper limit since $\mathcal{V}(\eta > \eta_{sr}) =0$, yields that  during the slow roll stage when $\mathcal{V}(\eta) =0$
   \be B_k(\eta   ) \simeq -i \frac{e^{-i\pi \nu}}{4k^2} \int^{\eta}_{-\infty}\,e^{-2ik\eta'}
   \mathcal{V}^{\,'}(\eta')   \, d\eta' \, \rightarrow |B_k(\eta)|^2 \lesssim \frac{1}{16 k^4} \,. \label{bofketaC}\ee This implies that for subhorizon modes
   \be |A_k(0)|^2 -1 = |B_k(0)|^2 \lesssim \frac{1}{k^4}\,,   \label{largk}\ee therefore the number of Bunch-Davies particles falls off very fast at large (subhorizon) momenta and the general initial conditions do not affect the short distance and renormalization aspects. Therefore, for modes that are deep within the Hubble radius during most of the slow roll era, and therefore, where very deep inside the Hubble radius during the pre-slow roll era it follows that
   \be \mathcal{T}(k)= 1 + \mathcal{O}(1/k^4) +\cdots \,.\label{toflargek}\ee

   Although the intermediate range of momenta must be studied numerically for definite realization of the pre-slow roll potentials there are several relevant consequences of the results obtained in the superhorizon and subhorizon limits:

   \begin{itemize}
   \item{On the largest scales \emph{today} corresponding to wavevectors that crossed the horizon $\sim 60$   e-folds before the end of inflation, the initial conditions set by a pre-slow roll rapid dynamical evolution of the inflaton yields a \emph{suppression} of the power spectrum when the potential $\mathcal{V}(\eta)$ is attractive, this is the situation for  a ``fast-roll'' stage as confirmed numerically in refs. \cite{boyan3,hectordestri,reviunos,lellor}. This suppression \emph{may} explain at least the large scale anomaly in the CMB reflected on the low power for the lowest multipoles\footnote{Although it is unlikely to explain the low multipole alignment or large scale asymmetry.}.  }

   \item{The effect of pre-slow roll initial conditions is negligible on small scales, those that crossed the horizon late or near the end of slow roll inflation. For example scales that reentered at the time of recombination imprinted on the first acoustic peaks, crossed out during  $\simeq 10$ e-folds in the period lasting about $60$ e-folds before the end of inflation. These modes were deep inside the Hubble radius during the pre-slow roll stage ($\gtrsim 60$ e-folds prior to the end of inflation) and their contribution to $\mathcal{T}(k)$ is strongly suppressed.

       This suggests that   these initial conditions \emph{may} suppress the power spectrum for  the largest scales but do not modify the spectral index and do not introduce a significant running of the spectral index with wavevector. }

   \end{itemize}

   Although this latter consequence must be studied in further detail numerically, we now focus on the impact of these type of initial conditions upon the infrared aspects of correlations for light scalar fields during de Sitter inflation, postponing a detailed analysis for curvature perturbations to further study. In particular, we have found that whereas individually the Bogoliubov coefficients feature large contributions for superhorizon momenta (as determined by the result for the sum $A_k+B_k \propto   k^{-2\nu}$), the power spectrum is only sensitive to the \emph{difference} and is smooth with a finite limit for superhorizon momenta, thus the question remains: are there any other infrared sensitive quantities that may feature a stronger dependence on initial conditions?. We study below the following infrared aspects: the self-consistent generation of mass and the decay width of single particle states during de Sitter inflation, both are consequences of a strong infrared enhancement of nearly massless fields in inflationary cosmology, and cross-correlation between sub and superhorizon modes in the decay products.

\section{Infrared aspects of scalar field correlations. }

Our goal is to study the influence of initial conditions on infrared aspects of scalar field correlators, in particular to assess how initial conditions arising from the pre-slow roll stage  modify the self-consistent mass generated by infrared fluctuations and also how they affect the decay of single particle states and cross-horizon correlations.

For the purposes of this work, only minimally coupled scalar field theories in a spatially flat de Sitter cosmology (the limit $\epsilon_V; \eta_V \rightarrow 0$) will be considered. The action for this field is given by

\be
I = \int d^3x \, dt \, a^3(t) \left\{ \frac{1}{2} \dot{\phi}^2 - \frac{(\nabla \phi)^2}{2a^2} - V(\phi) \right\}
\ee The potential under consideration will be of the form

\be
V(\phi) = \frac{1}{2}  M^2   \phi^2 + \lambda \phi^p ~~;~~ p=3,4
\ee   Passing to conformal time and conformally rescaling the fields

\be
a(t(\eta)) \equiv C(\eta) = \frac{-1}{H \eta} ~~;~~ a(t) \phi(\vx,t) \equiv \chi(\vx,\eta)\,,
\ee   the action can be rewritten, after discarding surface terms, as

\bea
&& I = \int d^3x \, d \eta \left\{ \frac{1}{2} \left[ \chi'^2 - (\nabla \chi)^2 - \mathcal{M}^2(\eta) \chi^2 \right] - \lambda (C(\eta))^{4-p} \chi^p \right\} \\
&& \mathcal{M}^2(\eta) \equiv  M^2   C^2(\eta) - \frac{C''(\eta)}{C(\eta)} =  \frac{1}{\eta^2}\left[ \frac{M^2}{H^2} -2 \right]
 \eea  where $' = d / d \eta$. The equations of motion for the Fourier modes in the non-interacting theory during the de Sitter stage become

\be
\chi''_{\vk} ( \eta) + \left[k^2 - \frac{1}{\eta^2} \left( \nu^2 - \frac{1}{4} \right) \right] \chi_{\vk} (\eta) = 0 ~~;~~ \nu^2 = \frac{9}{4} - \frac{M^2}{H^2}
\ee

Furthermore, we focus on light, nearly massless fields with $M^2/H^2 \ll 1$  in exact de Sitter space time in which case it follows that $\epsilon_V = \eta_V =0$ and
\be \nu = \frac{3}{2} - \Delta ~~;~~\Delta = \frac{M^2}{3H^2} +\cdots \ll 1\,.  \label{deltadef}\ee
Infrared divergences arising from the nearly masslessness of the fields are manifest as poles in $\Delta$ in the various correlation functions\cite{holmanburgess,rigo,garb,serreau2,boyquasi,riottosloth}, we will focus on the leading order infrared contributions arising from the poles in $\Delta$.

In order to study the effect of initial conditions set by  a pre-de Sitter stage, we now quantize the scalar field with the general mode functions (\ref{Ssr}),

\be
\chi (\eta, x) =  \frac{1}{\sqrt{V}} \sum_{k}  \alpha_{k} S_{\nu}(k, \eta) e^{i\vec{k} \cdot \vec{x}} + \alpha^{\dagger}_k S^{*}_{\nu}(k, \eta) e^{-i\vec{k} \cdot \vec{x}} \label{field}
\ee where $S_1 = A_k g_{\nu}(k, \eta) + B_k g^{*}_{\nu}(k, \eta)$ and $\alpha |0_{\alpha} \rangle = 0$ defines the   vacuum with general initial conditions and the Bunch-Davies mode functions are given by (\ref{BDmodes}), and the coefficients $A_k,B_k$ obey the relation (\ref{constraint}).

Two results obtained in the previous section are relevant for the analysis that follows:
\bea && \mathcal{T}(k)    =    |A_k - B_k|^2 ~~\overrightarrow{k\rightarrow 0} ~~  \mathcal{T}(0)+\mathcal{O}(k^2)+\cdots \label{limit}\\ &&  |A_k| ~~\overrightarrow{k\rightarrow \infty} ~~ 1 + \mathcal{O}(1/k^4)~~;~~ |B_k| ~~\overrightarrow{k\rightarrow \infty} ~~ \mathcal{O}(1/k^2) \label{limAB} \eea  With $\mathcal{T}(k)$ a smooth function of $k$ and $\mathcal{T}(0)$ given by (\ref{finiT0}).

\subsection{Interaction Picture}

The time evolution of interacting fields is handled in a straightforward manner. In the Schrodinger picture, a quantum state $|\Psi (\eta) \rangle$ obeys

\be
i \frac{d}{d\eta} | \Psi (\eta) \rangle = H(\eta) | \Psi (\eta) \rangle
\ee where the Hamiltonian $H(\eta)$ is explicitly a function of $\eta$ in an expanding cosmology. Defining the time evolution operator, this has the formal solution

\be
i \frac{d}{d\eta} U(\eta, \eta_0) = H(\eta) U(\eta, \eta_0) ~~;~~ U(\eta_0, \eta_0) = 1
\ee so that $|\Psi (\eta) \rangle = U(\eta, \eta_0) | \Psi (\eta_0) \rangle$. The Hamiltonian can be separated into free and interacting pieces, $H(\eta) = H_0 (\eta) + H_i(\eta)$, where $H_0$ is the  non-interaction Hamiltonian. Defining the time evolution operator for the free theory, $U_0 (\eta, \eta_0)$, so that

\be
i \frac{d}{d\eta} U_0(\eta, \eta_0) = H_0(\eta) U_0(\eta, \eta_0) ~~;~~ i \frac{d}{d\eta} U^{-1}_0(\eta, \eta_0) = -  U^{-1}_0(\eta, \eta_0) H_0(\eta)  ~~;~~ U(\eta_0, \eta_0) = 1
\ee From here, the interaction picture may be defined in the usual manner as

\be
|\Psi(\eta) \rangle_I = U_I (\eta, \eta_0) |\Psi(\eta_0) \rangle_I = U_0^{-1} (\eta, \eta_0) |\Psi (\eta) \rangle
\ee so that $U_I(\eta, \eta_0)$ is the interaction picture time evolution operator such that

\be
\frac{d}{d\eta} U_I(\eta, \eta_0) = -iH_I(\eta) U_I(\eta, \eta_0) ~~;~~ U_I(\eta_0, \eta_0) =1 ~~;~~ H_I(\eta) = U^{-1}_0 (\eta, \eta_0) H_i(\eta) U_0(\eta,\eta_0)
\ee For the interactions that will be considered here, the interaction Hamiltonian is given explicitly by

\be
H_I (\eta) = \frac{\lambda}{(-H \eta)^{4-p}} \int d^3x (\chi(\vx, \eta))^p
\ee To leading order in $\lambda$, the standard solution in perturbation theory is

\be
U_I(\eta, \eta_0) = 1 -i \int^{\eta}_{\eta_0} d\eta' H_I(\eta') +...
\ee

\subsection{The infrared contribution to the tadpole: }

The tadpole, $\langle 0 | \chi^2(\vec{x},\eta) | 0 \rangle$ with $|0\rangle$ being the vacuum with non-Bunch Davies initial conditions, will play an important role in the following discussion.   It is given by

\be \langle 0 | \, \chi^2(\vx, \eta) |\, 0 \rangle = \int \frac{d^3k}{(2 \pi)^{\,3}} | \, S(k, \eta)|^2  \,. \label{tadpi}\ee   Our goal is to extract the most relevant infrared contributions. In order to understand the influence of the Bogoliubov coefficients $A_k;B_k$ determined by the initial conditions, we revisit the evaluation of the tadpole   for the Bunch-Davies case, namely $A_k=1;B_k=0$, $S(k;\eta)=g_\nu(k;\eta)$  to highlight the origin of the most infrared relevant contributions. In this case  making a change of variables $y = -k\eta$ the tadpole is given by
\be {}_{BD}\langle 0 | \, \chi^2(\vx, \eta) |\, 0 \rangle_{BD} = \frac{1}{8\pi\,\eta^2} \int^{\Lambda_p/H}_0 \frac{dy}{y}\,y^3 |H^{(1)}_\nu(y)|^2 \ee where we have introduced an ultraviolet cutoff in physical coordinates. To isolate the infrared divergences for $\Delta \ll 1$ we write the integral above as
\be \int^{\Lambda_p/H}_0 \frac{dy}{y}\,y^3 |H^{(1)}_\nu(y)|^2 = \int^{\mu_p/H}_0 \frac{dy}{y}\,y^3 |H^{(1)}_\nu(y)|^2 + \int^{\Lambda_p/H}_{\mu_p/H} \frac{dy}{y}\,y^3 |H^{(1)}_\nu(y)|^2 \label{intsplit}\ee with $\mu_p \rightarrow 0$ an infrared physical cutoff. For the first integral  we use $\nu = 3/2 - \Delta$ with $ 0< \Delta \ll 1$ and
\be  z^3 \, \left|H^{(1)}_\nu(z)\right|^2 \buildrel{z \to
0}\over=\left[ \frac{2^{\nu} \; \Gamma(\nu)}{\pi} \right]^2 \; z^{2
\, \Delta} \label{smallz}\ee thus $\Delta > 0$ regulates the infrared behavior of the tadpole and the first integral yields
\be\label{IRint} \int^{\frac{\mu_p}{H}}_0 \frac{dz}{z} \; z^3 \,
\left|H^{(1)}_\nu(z)\right|^2 = \frac{2}{\pi}\left[\frac{1}{2 \,
\Delta}+ \frac{ {\mu^2_p}}{2H^2} + \gamma - 2 + \ln   \frac{2 \mu_p}{H}
+\mathcal{O}(\Delta)\right]\,, \ee \noindent where we have displayed
the pole in $\Delta$ and the leading infrared logarithm. In the second integral in (\ref{intsplit}) we set $\nu =3/2$ and combining its result with (\ref{IRint} ) we find that  the dependence on the infrared cutoff $\mu_p$ cancels in the limit $\mu_p\rightarrow 0$ leading to the following   final result for the tadpole with Bunch-Davies vacuum
\be
 {}_{BD}\langle 0 | \, \chi^2(\vx, \eta) |\, 0 \rangle_{BD} = \frac{1}{8\pi^2 \,\eta^2}
\left[ \frac{{\Lambda_p}^2}{H^2} + 2 \ln \frac{\Lambda_p }{H}+\frac1{\Delta}
 + 2 \, \gamma - 4 + \mathcal{O}(\Delta) \right]\,, \label{BDtad}
\ee  While the
quadratic and logarithmic \emph{ultraviolet} divergences are
regularization scheme dependent, the pole in $\Delta$ arises from
the  infrared behavior and is independent of the regularization
scheme. In particular this pole coincides with that found
in the expression for $<\phi^2(\vx,t)>$ in refs.\cite{boyquasi,holmanburgess,rigo,smit}. The
\emph{ultraviolet divergences}, in whichever renormalization scheme,
 require that the effective field theory be
defined to contain \emph{renormalization counterterms} in the bare
effective Lagrangian, for the tadpole this counterterm is of the form $\chi(\eta)\,J(\eta)$  and $J(n)$ is required to cancel    the ultraviolet divergences. Thus,   the { \it{renormalized}} tadpole in the Bunch-Davies vacuum is given by
\be \mathcal{I}_{BD}(\eta)\equiv  {}_{BD}\langle 0 | \, \chi^2(\vx, \eta) |\, 0 \rangle^{ren}_{BD} = \frac{1}{8\pi^2 \,\eta^2}~\frac1{\Delta} ~\big[1+
  \cdots \big]\,, \label{poletad}
\ee
\noindent where the dots stand for higher order terms in $\Delta \ll 1$.

We are now in position to understand the effect of non-Bunch-Davies initial conditions. The most infrared divergent contribution is determined by superhorizon modes for which $g_\nu(k;\eta) \simeq i\sqrt{-\pi \eta}\,Y_\nu(-k\eta)/2$ hence
\be |S(k;\eta)|^2 \simeq \frac{-\pi \eta}{4}  Y^2_\nu(-k\eta)\,\mathcal{T}(k) ~~;~~ -k\eta \ll 1 \label{sups}\ee the fast fall off of the Bogoliubov coefficients with large $k$ entails that the ultraviolet behavior of the tadpole is the same as in Bunch-Davies vacuum so that renormalization of the tadpole proceeds just as in the Bunch-Davies case. The pole in $\Delta$ in (\ref{BDtad}) arises from a narrow band of superhorizon wavevectors with the infrared cutoff $\mu \rightarrow 0$. The results of the previous section show that for superhorizon wavevectors $\mathcal{T}(k)= \mathcal{T}(0)+\mathcal{O}(k^2)+\cdots$ is a smooth function of $k$ with $\mathcal{T}(0)$ given by (\ref{finiT0}). Therefore, to obtain the leading order infrared contribution for $\Delta \ll 1$ we replace $\mathcal{T}(k) \rightarrow \mathcal{T}(0)$ in (\ref{sups}) because the higher powers of $k$ in  $\mathcal{T}(k)$ yield terms that are subleading for $\Delta \ll 1$. Furthermore since for large $k$ we found that $\mathcal{T}(k) \lesssim 1/k^4$ the ultraviolet divergences of the tadpole are the same as for the Bunch-Davies case and renormalization is achieved in the same manner as with Bunch-Davies initial conditions.

Therefore for general initial conditions set during a pre-slow roll stage we obtain
\be \mathcal{I}(\eta)\equiv   \langle 0 | \, \chi^2(\vx, \eta) |\, 0 \rangle^{ren}  = \frac{1}{8\pi^2 \,\eta^2}~\frac{\mathcal{T}(0)}{\Delta} ~\big[1+
  \cdots \big]\,, \label{poletadgen}
\ee

Although this discussion was focused on the tadpole, similar arguments will allow to extract the leading infrared contributions in other correlators, the main point is that the leading infrared divergences that are responsible for poles in $\Delta \ll 1$ arise from a small band of superhorizon wavevectors for which $\mathcal{T}(k) \simeq \mathcal{T}(0)$.

\subsection{Self Consistent Mass Generation}

\subsubsection{ $\phi^3$ theory}

For this particular field theory, radiative corrections will induce a non zero expectation value of the field in the "dressed" vacuum. At leading order for a general interaction Hamiltonian, the dressed vacuum evolves in time as

\be
|\tilde{0} (\eta) \rangle = U(\eta, \eta_o) |\tilde{0} (\eta_o)\rangle \simeq \left(1 - i\int^{\eta}_{\eta_o} d\eta^{'} H_I (\eta^{'}) +...\right) |\tilde{0} (\eta_o)\rangle
\ee so that, to leading order, the expectation value of the field is given by

\be
\langle \tilde{0} (\eta) | \, \chi(y,\eta) | \, \tilde{0} (\eta) \rangle - \langle \tilde{0} (\eta_o) |\, \chi(y,\eta) |\, \tilde{0} (\eta_o) \rangle \equiv \delta \langle \chi(y,\eta) \rangle = i \langle \tilde{0} (\eta_o) | \int^{\eta}_{\eta_o} d\eta^{'} \left[ H_I (\eta'), \chi(y,\eta) \right] | \tilde{0} (\eta_o) \rangle
\ee   Specializing to $\lambda \phi^3$ theory results in

\be
\delta \langle \chi \rangle = 3i \lambda \int^{\eta}_{\eta_o} d\eta\,' C(\eta') \int d^3x [\chi(x, \eta\,'), \chi(y, \eta) ] \langle \tilde{0} (\eta_o) | \chi^2(x, \eta') | \tilde{0} (\eta_o) \rangle
\ee where the commutator is readily evaluated using the expansion of the field and creation/annhilation operator commutation relations, the result being

\be
\int d^3x [\chi(x, \eta'), \chi(y, \eta) ] = \left[ S(k;\eta') S^{*}(k; \eta) - S^{*}(k ;\eta') S(k ; \eta)\right]_{k=0}
\ee This is readily evaluated using the limiting form of Bessel functions and it can be shown that

\be
\begin{split}
& S(k,\eta')S^{*}(k, \eta)\Big|_{k\rightarrow 0} = \\
& \frac{-\pi \sqrt{\eta \eta'}}{4} \Bigg\{ (|A_k|^2 + |B_k|^2) \left(\frac{1}{\Gamma^2(\nu +1)} \left( \frac{k \eta \eta'}{4} \right)^{\nu} + \frac{\Gamma^2(\nu)}{\pi^2} \left(\frac{k \eta \eta'}{4}\right)^{- \nu} \right) + \\
& \left(A_kB^{*}_k  + B_kA^{*}_k  \right) \left( \frac{1}{\Gamma^2(\nu+1)} \left(\frac{k^2 \eta \eta'}{4}\right) -\frac{\Gamma^2(\nu)}{\pi^2} \left(\frac{k \eta \eta'}{4} \right)^{-\nu} \right)+  \\
& \left(A_kB^{*}_k   - B_k A^{*}_k  \right) \left(\frac{-i}{\pi \nu} \right) \left( \left( \frac{\eta}{\eta'}^{-\nu}\right) + \left( \frac{\eta'}{\eta}^{-\nu} \right) \right) \\
& (|A_k|^2 - |B_k|^2) \left(\frac{-i}{\pi \nu}\right) \left[ \left(\frac{\eta'}{\eta}\right)^{-\nu} - \left(\frac{\eta}{\eta'}\right)^{-\nu} \right] \Bigg\}\,,
\end{split}
\ee note that the first three terms would diverge in the long wavelength limit, however these are all \emph{real}, and  $S(k,\eta')S^{*}(k,\eta) - S^{*}(k, \eta')S(k,\eta) = 2i \,\mathrm{Im} (S(k, \eta') S^{*} (k,\eta))$, hence these terms cancel in the expectation value. Since  $|A_k|^2-|B_k|^2=1$, the commutator becomes

\be
\int d^3x [\chi(x, \eta'), \chi(y, \eta) ] = \frac{i}{2\nu} (\eta^{\beta_+} \eta'^{\beta_-} - \eta^{\beta_-} \eta'^{\beta_+}) ~~;~~ \beta_{\pm} = \frac{1}{2} \pm \nu
\ee which is independent of the  vacuum state.

Therefore, the full expression for the expectation value becomes

\be
\delta \langle \chi \rangle = \frac{-3 \lambda}{2 \nu H} \int^{\eta}_{\eta_o} \frac{d\eta'}{\eta'} \left[ \eta^{\beta_+} \eta'^{\beta_-} - \eta^{\beta_-} \eta'^{\beta_+} \right] \langle 0 | \, \chi^2(\vx, \eta') |\, 0 \rangle \label{expect}
\ee To leading order in $\Delta$ the renormalized tadpole contribution is given by (\ref{poletadgen})

\be
\delta \langle \chi \rangle = \frac{-3 \lambda\,\mathcal{T}(0)}{16 \pi \nu H\Delta} \int^{\eta}_{\eta_o} \frac{d\eta'}{\eta'^3} \left[ \eta^{\beta_+} \eta'^{\beta_-} - \eta^{\beta_-} \eta'^{\beta_+} \right] = \frac{- \lambda\,\mathcal{T}(0)}{8 \pi \Delta H \eta} \left( \frac{1}{\Delta} \left( 1 - \left(\frac{\eta}{\eta_o} \right)^{\Delta} \right) - \frac{1}{3} \left( 1 - \frac{\eta^3}{\eta_o^3}  \right) \right)  \ee  therefore to leading order in $\Delta$ and as $\eta/\eta_0 \rightarrow 0$ we find

\be
\delta \langle \chi \rangle = \frac{- \lambda \mathcal{T}(0)}{8 \pi H \Delta^2 \eta}  + \mathcal{O}(\Delta)
\ee

If   the field initially has vanishing  expectation value the interactions lead to a non-vanishing expectation value in the interacting ground state asymptotically given by

\be
\langle 0| \chi(y, \eta)|0 \rangle = \bar{\chi} (\eta) \rightarrow \frac{ - \lambda}{8 \pi^2 H \eta} \frac{\mathcal{T}(0)}{\Delta^2}\ + \mathcal{O}(\Delta) \label{chiexp}
\ee Then the \emph{unscaled} field obtains a constant expectation value for $\eta/\eta_o \rightarrow 0$,

\be
\langle \phi(y, \eta) \rangle = \frac{1}{a(\eta)} \langle \chi(y, \eta) \rangle = \frac{\lambda}{8 \pi^2} \frac{\mathcal{T}(0)}{\Delta^2}\ + \mathcal{O}(\Delta) \,.
\ee
This result which includes the effect of initial conditions is a generalization of that found in ref.\cite{boyquasi} and is noteworthy because infrared effects lead to an asymptotic expectation value which is time independent, signaling the emergence of a non-trivial minimum of an effective action.

The emergence of a non-trivial expectation value and a mininum of the effective action implies that it is necessary to redefine the field shifting by this expectation value, namely

\be
\chi(x, \eta) = \Psi(x, \eta) + \bar{\chi} (\eta) ~~;~~ \langle \tilde{0} | \Psi (x, \eta) | \tilde{0} \rangle = 0
\ee

This is the origin of the mechanism of self-consistent mass generation, for consider that the bare Lagrangian describes a massless scalar field with cubic interaction, shifting by the vacuum expectation value, the cubic term now written in terms of $\Psi$ becomes

\be
H_I = \int d^3x \left[ \frac{1}{\eta^2} \frac{M^2}{2H^2} \Psi^2 - \frac{\lambda}{H \eta} \Psi^3 \right] \label{shiftedham}
\ee where
\be \frac{1}{\eta^2} \frac{M^2}{2 H^2} = -3 \frac{\lambda}{H \eta} \bar{\chi} ( \eta)\,.  \label{masss}\ee This suggests a \emph{self-consistent}  mass generation mechanism by replacing $\bar{\chi}$ by the result (\ref{chiexp}), namely

\be
\frac{1}{\eta^2} \frac{M^2}{2 H^2} =  \frac{3 \lambda^2}{8 \pi^2 H^2 \eta^2}   \frac{\mathcal{T}(0)}{\Delta^2}\ \Big[1 + \mathcal{O}(\Delta)\Big] \label{scsol}
\ee since $\Delta = M^2/3H^2$ this is a self consistent condition with the result

\be
M = H \sqrt{3} \left( \frac{\lambda}{2 \pi H} \right)^{1/3} \Big[\mathcal{T}(0) \Big]^{1/6}\equiv M_{BD}\,\Big[\mathcal{T}(0) \Big]^{1/6}\label{scM}
\ee where $M_{BD}$ is the self-consistent mass obtained with Bunch-Davies initial conditions\cite{boyquasi}. It is reassuring to find that the sign of the induced expectation value is consistent with $M^2 >0$, otherwise the radiatively induced squared mass would indicate an instability in the theory.

 This is a noteworthy result, the strong infrared behavior leads to a self-consistent mass generation which is \emph{non-analytic} in the transfer function for initial conditions.

\subsubsection{$\phi^4$ theory}

For this theory, the Lagrangian density is now $\mathcal{L}_I = -\lambda \chi^4$ and, as discussed previously, the expectation value of the field remains zero. As discussed in ref.\cite{boyquasi} the mechanism of self consistent mass generation for a massless field is accomplished   by introducing a mass term in the free Lagrangian and then subtracting it out again as a counterterm in the interaction part
 \be
\mathcal{L}_I = \frac{1}{2} C^2(\eta) M^2 \chi^2 - \lambda \chi^4
\ee
 and requesting that the tadpole cancels the mass counterterm leading to a self-consistent condition akin to the Hartree resummation\cite{garb,rigo,arai,serreau2}, namely
\be
\frac{M^2}{2 H^2 \eta^2} = 6 \lambda \,\langle 0| \chi^2(x,\eta)|0 \rangle
\ee  where the renormalized tadpole is given by (\ref{poletadgen}), therefore to leading order in $\Delta$, the self consistent mass becomes

\be
M = H \left[ \frac{9 \lambda\,\mathcal{T}(0)}{2 \pi^2}  \right]^{1/4} \equiv M_{BD}\, \left[ \mathcal{T}(0)   \right]^{1/4}\,. \label{scmassfi4}
\ee

Again the Bunch-Davies case corresponds to $\mathcal{T}(0)=1$ thus the self-consistent condition leading to dynamical mass generation from infrared divergences yields a non-analytic dependence of the generated mass upon the initial conditions.

The comparison between the infrared generated mass for Bunch-Davies initial conditions and the puzzling discrepancy  obtained with other approaches\cite{staro1,rajaraman,holmanburgess,rigo,garb,serreau2} has been discussed in ref.\cite{boyquasi} (see the first reference).

\subsection{Initial condition dependent anomalous dimensions:}

The self-consistent mass generation through infrared divergences lead to the following expressions for $\Delta$ from the self-consistent solutions for cubic $(3)$  and quartic $(4)$ interactions respectively,
\bea \Delta_{(3)} & = &  \Bigg[ \frac{\lambda_{(3)}\,\sqrt{\mathcal{T}(0)}}{2\pi\,H}\Bigg]^\frac{2}{3}\,\,, \label{delfi3} \\ \Delta_{(4)} & = & \Big[\frac{\lambda_{(4)} \, \mathcal{T}(0)}{2\pi^2} \Big]^{\frac{1}{2}} \,. \label{delfi4}
\eea where $\lambda_{(3)},\lambda_{(4)}$ are the cubic and quartic couplings respectively.

 This result, in turn, implies that the power spectrum acquires non-perturbative initial condition-dependent \emph{anomalous dimensions}, namely

\be \mathcal{P} \propto k^3 \langle 0|\chi_{\vec{k}}(\eta)\, \chi_{-\vec{k}}(\eta)|0\rangle \propto k^{2\Delta}\,. \label{anodimself}\ee  where $\Delta$ is given by (\ref{delfi3},\ref{delfi4}) for cubic and quartic self-interactions respectively.

We highlight that for initial conditions determined by a fast-roll stage prior to slow roll, the long-wavelength power spectrum is suppressed and all the corrections from the initial conditions on self-consistent masses and anomalous dimensions are suppressed with respect to the Bunch-Davis result. Hence, initial conditions that \emph{could} explain the anomalously low quadrupole in the CMB lead consistently to a suppression of all infrared effects, including the non-perturbatively generated masses and anomalous dimensions.

\section{Particle Decay: width dependence on initial conditions. }

In an expanding cosmology, the lack of a global time-like Killing vector implies the lack of thresholds for particle decay (a consequence of energy-momentum conservation). Therefore, a single particle state of a field can decay into multiple particle states of the \emph{same field} as discussed in refs.\cite{boyprem,boyan} confirmed for heavy fields in ref.\cite{moschella,leblond} and more generally (and thoroughly) for a scalar theory with cubic interactions in\cite{marolf}.  The usual method to extract a decay rate in Minkowski space-time relies on energy-momentum conservation that leads to a transition probability that grows linearly in time at long times, namely a time-independent \emph{decay rate}. The lack of energy conservation in an expanding cosmology prevents the usual implementation of what is, essentially, Fermi's Golden rule, instead the transition probability and ultimately the full time evolution of quantum states must be studied in detail.

 In ref.\cite{boyholds} a non-perturbative field theoretical generalization of the Wigner-Weisskopf method to study the decay of single particle states was adapted to inflationary cosmology, and in ref.\cite{boyquasi} this method was generalized and extended to obtain in a consistent manner both the infrared induced self-consistent masses and the time dependent  decay   width of  particle states.  The details of these methods have been explained in detail in refs.\cite{boyholds,boyquasi,lellosup,lellomink} and the reader is referred to these references for details. For self-consistency we give a brief summary of the method in appendix (\ref{appb}).

\subsection{Transition Amplitude and Probability: cubic vertex} \label{decaySec}

To identify the corrections to masses and the decay widths, consider the interaction of a scalar fields through a cubic vertex. The interaction Hamiltonian is given by

\be
H_i = \lambda \int d^3x \, a(t)^3   \phi^3 = \lambda C(\eta) \int d^3x \, \chi^3(x,\eta)
\ee where the conformally rescaled fields have been used.  Using the expansion of the field, Eq (\ref{field}), the matrix element for process $\chi \rightarrow 2 \chi$ can be readily obtained, it is given by

\be
\mathcal{A_{\chi \rightarrow \chi \chi}} = \frac{-6i\lambda}{V^{1/2}} \int^{\eta}_{\eta_o} d \eta' C(\eta') \, S(k,\eta') S^*(k-q,\eta') S^*(q,\eta')\,, \label{amplit}
\ee and the  total transition probability   is  given by

\be
P_{\chi \rightarrow \chi \chi} = \sum_q |\mathcal{A}|^2 \equiv \int^{\eta}_{\eta_o} d\eta_1 \, d\eta_2 \, \Sigma (k, \eta_1, \eta_2)
\ee where

\be
\Sigma (k, \eta_1, \eta_2) = \frac{ 36 \lambda^2}{H^2 \eta_1 \eta_2} \int \frac{d^3q}{(2\pi)^3} \,   S^*(k,\eta_1) S(k,\eta_2) S(k-q,\eta_1) S^*(k-q,\eta_2) S(q,\eta_1) S^*(q,\eta_2) \,,\label{sigma}
\ee with the property

\be
\Sigma (k, \eta_1, \eta_2) = \Sigma^* (k, \eta_2, \eta_1) \label{complex}
\ee Inserting a factor of $1= \theta(\eta_2 -\eta_1) + \theta(\eta_1 -\eta_2)$ in the integral and making use of (\ref{complex}) yields

\be
P_{\chi \rightarrow \chi \chi}(k,\eta) = 2\int^{\eta}_{\eta_0} d\eta_2 \int^{\eta_2}_{\eta_0} d\eta_1 \mathrm{Re}\, \left[ \Sigma (k, \eta_1, \eta_2) \right]
\ee so the \emph{transition rate} is easily identified   to be

\be
\Gamma(\eta) = \frac{d}{d\eta} P_{\chi \rightarrow \chi \chi}(k,\eta) = 2 \int^{\eta}_{\eta_0} d \eta' \, \mathrm{Re}  \left[ \Sigma (k, \eta, \eta') \right]\,. \label{decay}
\ee

In Minkowski space time where energy-momentum conservation holds, the transition probability for a decaying state grows linearly (secularly) in time leading to a constant transition \emph{rate} and an overall energy momentum delta function in the phase space integrals determining the kinematic reaction thresholds. Only when the transition probability grows with time is the process associated with the decay of the parent particle.

In an expanding cosmology there lack of energy conservation (energy momentum is covariantly conserved) leads to the lack of kinematic thresholds and the decay process $\chi \rightarrow 2\chi$ is allowed\cite{boyan,boyholds}. In ref.\cite{boyquasi} it is shown in detail non-perturbatively that  an initial single particle state decays as
\be  |\Psi(\eta)\rangle \propto |\Psi(\eta_0)\rangle~e^{-\frac{1}{2}\int^\eta_{\eta_0} \Gamma(\eta')\,d\eta'}\,. \label{statedecay}\ee

\subsection{  Decay Rate}

In order to calculate the decay rate of $\chi \rightarrow 2\chi$ we need to evaluate   $\Sigma(k,\eta_1,\eta_2)$  given by eqn.  (\ref{sigma}). We focus on the long time limit $\eta_1,\eta_2 \rightarrow 0$ and the leading order in $\Delta$. The calculation  is involved and has been carried out in detail for the case of Bunch-Davies initial conditions in ref.\cite{boyquasi}, the details  of this calculation for general initial conditions with the Bogoliubov coefficients are relegated to   appendix (\ref{appA}).

 We find to leading order in $\Delta$ and in the long time limit,
\be
\Sigma(k,\eta_1,\eta_2) = \frac{18\,\lambda^2\,\mathcal{T}(0)}{\pi^2\,H^2\,\Delta   } ~  \frac{\left| S(k, \eta_1)\right|^2}{(\eta_1)^2}\frac{ \left| S(k, \eta_2)\right|^2}{(  \eta_2)^2}\,    + \mathcal{O}(\Delta^0)
\ee

The factor $\mathcal{T}(0)$ originates in the infrared region that yields the pole in $\Delta$ corresponding to one of the internal lines  in the self energy,  either $q\simeq 0$ or $q\simeq k$,  within the band of superhorizon wavevectors. To leading order in $\Delta$ the self energy is purely real and  the decay rate becomes

\be
\Gamma(k;\eta) = \frac{36\lambda^2}{\pi^2 H^2 } \frac{\mathcal{T}(0)}{\Delta} \frac{|S(k,\eta)|^2}{\eta^2}   \int_{-\eta }^{-\eta_0} d(-\eta')  \frac{|S(k,\eta')|^2}{(\eta')^2}  \label{gamaf}
\ee

At long times when the external momentum $k$ crosses the Hubble radius, this expression simplifies a few-efolds after crossing since in this limit $|S(k;\eta)|^2 \rightarrow \mathcal{T}(k)\,(-\pi \eta/4)\,Y^2_\nu(-k\eta)$ and using the expression (\ref{suphorbes}) we find in this limit
\be \Gamma(k;\eta) \simeq \frac{9\lambda^2 \mathcal{T}(0)\mathcal{T}^2(k)}{\pi^2H^2\Delta (-\eta)(-k\eta)^6} \label{gamasuphori}\ee The Bunch-Davies result is obtained by replacing $\mathcal{T}(k) \rightarrow 1$ and coincides with the result obtained in ref.\cite{boyquasi}\footnote{There is a factor 2 error in the prefactor in this reference.}.

\textbf{  Simple rules:}

The analysis presented above yields as corollary the following  set of simple rules to assess the effect of non-Bunch-Davies in the correlators:

\begin{itemize}
\item{ Correlation functions feature products of mode functions of the form $S(k,\eta)S^*(k,\eta')$, for values of $k$ so that $-k\eta, -k\eta' \gg 1$ this product can be replaced by
    \be S(k,\eta)S^*(k,\eta') \rightarrow \frac{\pi}{4}\, \mathcal{T}(k) \,  \big( \eta\,\eta'\big)^{1/2}\,Y_\nu(-k\eta)Y_\nu(-k\eta')\,. \label{replace}\ee }

\item{In the momentum integrals that lead to infrared divergences and resulting in poles in $\Delta$, the initial condition transfer function can be expanded as $\mathcal{T}(k) \simeq \mathcal{T}(0) + \mathcal{O}(k^2)+\cdots$, the higher order powers of $k$ do not yield infrared enhancements, therefore the poles in $\Delta$ are multiplied by $\mathcal{T}(0)$. Namely for poles in $\Delta$ that arise from momentum integration it follows that
    \be \frac{1}{\Delta} \rightarrow \frac{\mathcal{T}(0)}{\Delta}\,. \label{replacepoles}\ee }

\end{itemize}
These simple rules allow to extract the contribution from non-Bunch-Davies initial conditions, encoded in $\mathcal{T}$ to the various correlation functions.

\section{Entanglement Entropy: effect of initial conditions on correlations across the horizon}

In the  $\lambda \phi^3$ theory considered here, a single particle state, $|1_{\vk}\rangle$,   decays into a two particle state, $|1_{\vk-\vp}\rangle |1_{\vp}\rangle$ with the corresponding amplitude given by (\ref{amplit}). The full quantum state obtained from the time evolution is a linear superposition of the two particle states summed over the momentum $\vp$. Such a quantum state is \emph{entangled}. This is a general result highlighted in ref.\cite{lellomink}: the decay of a single particle state leads to a \emph{quantum entangled state} with correlations between the daughter particles as a consequence of conservation laws. In a spatially flat Friedmann-Robertson-Walker cosmology spatial momentum is conserved. In ref.\cite{lellosup} it was realized that the decay of an initial single particle state with wavelength deep inside the Hubble radius produces two particle states which in the case of light fields the leading contribution in $\Delta$ corresponds to the decay into a subhorizon and a superhorizon particle. This is an entangled state with correlations between the daughter particles \emph{across the Hubble radius}. As discussed in detail in ref.\cite{lellosup} this process is dominated by the emission and absorption of superhorizon quanta, and therefore it is enhanced in the infrared by poles in $\Delta$ which is a hallmark of the infrared aspects associated with light fields in de Sitter (or near de Sitter) space time.

The main tool to study the time evolution of single particle states and the correlated quantum state resulting from the decay is the quantum field theory version of the Weisskopf-Wigner method introduced in refs.\cite{boyholds,boyquasi,lellosup,lellomink} where the reader is referred to for a detailed treatment, a brief description is included in appendix (\ref{appb}) for consistency.

 Considering an initial state $|1_{\vk}\rangle$ at initial time $\eta_0$ results in the following quantum state

\be
| \Psi(\eta) \rangle_I = C_{k}(\eta) |1_{\vk}\rangle + \sum_{\vp} C_p(k,\eta) |1_{\vk-\vp}\rangle |1_{\vp}\rangle \label{psi}
\ee where the coefficients $C_k,C_p$ are obtained through (\ref{wwcoeff}) and (\ref{initialcoeff}). It has been shown that the Wigner Weisskopf truncation is fully consistent with unitarity as shown in ref. \cite{lellosup}. For completeness, this is shown explicitly in appendix \ref{appb}.

 With a fully unitary prescription to obtain the coefficients, the pure state density matrix corresponding to the entangled state of eq.(\ref{psi}) may be written

\be
\rho(\eta) = |\Psi(\eta)\rangle \langle \Psi(\eta)|. \label{rho}
\ee Considering the situation where a subhorizon mode ($ \vk \gtrsim (-1/\eta)$) decays, tracing out superhorizon ($ \vp \lesssim (-1/\eta)$) modes leads to the \emph{mixed state} density matrix for modes whose wavelengths are \emph{inside} the horizon during the evolution. This is given by

\be
\rho_r(\eta) = |C_k(\eta)|^2 |1_{\vk}\rangle \langle 1_{\vk}| + 2 \sum_{p \lesssim (-1/\eta)} |C_{p}(k;\eta)|^2
|1_{\vk-\vp}\rangle \langle 1_{\vk-\vp}| \label{rhored}
\ee where the factor $2$ accounts for the two regions of superhorizon momenta $p < (-1/\eta)$ and $|\vk-\vp|< (-1/\eta)$ which yield the same contribution, as can be easily seen after a relabelling of momenta.

The entanglement entropy is given by the Von-Neumann entropy for the reduced density matrix, where one finds
\be
\mathcal{S}(\eta) = -n_k(\eta)\ln n_k(\eta) - 2\sum_{p \lesssim (-1/\eta)} n_p(\eta) \ln n_p(\eta) \label{entropy}
\ee where the occupation numbers of the initial and \emph{produced} quanta are given by

\be
n_k(\eta) = \langle \Psi(\eta)|a^\dagger_{\vk}\, a_{\vk}|\Psi(\eta)\rangle = |C_k(\eta)|^2,\ n_p(\eta) = \langle \Psi(\eta)|a^\dagger_{\vp}\, a_{\vp}|\Psi(\eta)\rangle = |C_p(k;\eta)|^2 \,.\label{ocupa}
\ee The unitarity relation from eq.(\ref{unimark}) implies that

\be
\sum_{\vp} n_p(\eta) = 1- n_k(\eta)\,. \label{unitanumber}
\ee as expected on physical grounds. At this point, all that remains to calculate the entropy for this process is a calculation of the coefficients, (\ref{wwcoeff}) and (\ref{initialcoeff}).

Using (\ref{sigma}), the coefficient \ref{initialcoeff} can be calculated. For $|\vp| \ll -1/\eta; |\vk|,|\vk-\vp| \gg -1/\eta$, the mode functions in \ref{field} reduce to

\be
S_{\nu}(k,\eta) \rightarrow \frac{1}{\sqrt{2k}} \Big[A_k\,e^{-ik\eta}+B_k \,e^{ ik\eta}\Big] ~~;~~ S_{\nu} (p, \eta) \rightarrow \frac{i}{\sqrt{2}} \frac{A_p - B_p}{(-\eta)^{1-\Delta} p^{3/2 - \Delta} } \label{limiting}
\ee

For momenta $k$ deep inside the Hubble radius the results (\ref{aofketaB},\ref{bofketaB},\ref{bofketaC}) justify to set $A_k =1~;~B_k=0$ to leading order. The integral in (\ref{sigma}) can be carried out with an infrared cutoff $\mu \lesssim (-1/\eta)$ and the leading order in $\Delta$ is extracted by approximating $\mathcal{T}(p) \simeq \mathcal{T}(0)$, leading to the result

\be
\Sigma(k,\eta_1,\eta_2) = \frac{\alpha}{k^{2-2\Delta} \eta_1^{2-\Delta} \eta_2^{2-\Delta}} ~~;~~  \label{sigfina} \ee where
\be \alpha \equiv \frac{9 \lambda^2 \mathcal{T}(0)}{8 \pi^2 H^2 \Delta} \,.\label{alfadef}\ee

Using this result, the coefficient of \ref{initialcoeff} becomes

\be
C_k (\eta) = \exp \left[-\frac{\alpha}{2z^{2-2\Delta}}\right] ~~;~~ z \equiv k\eta
\ee   The matrix element for this process is given by

\be
\begin{split}
& \mathcal{M}(p;k;\eta )= \langle 1_{\vk-\vp};1_{\vp}|H_I(\eta)|1_{\vk}\rangle = -\frac{6\lambda}{H\eta\sqrt{V}} \, S_\nu(k;\eta)\,S^*_{\nu }(p;\eta)\,S^*_{\nu }(|\vec{k}-\vec{p}|;\eta) \\
& \rightarrow \frac{-6\lambda(A_0^* - B_0^*)}{2\sqrt{2}kHV^{1/2}(-\eta)^{2-\Delta}p^{3/2-\Delta}} \label{mtxele1}
\end{split}
\ee so that

\be
C_{p}(k;\eta) = -i \int^{\eta}_{\eta_0} \mathcal{M}(p;k;\eta')\, C_k(\eta')\, d\eta'  = \frac{-6i\lambda(A_0^*-B_0^*)}{2\sqrt{2}H V^{1/2} p^{3/2-\Delta}}\frac{1}{\sqrt{\alpha}} \int_{y_0}^y e^{-y^2/2} dy \label{coefpk}
\ee where a change of variables, $\eta =\sqrt{\alpha} /ky$, has been made. In principle, this can be calculated in terms of error functions but unitarity provides a simpler means of evaluation. Since $\alpha \propto |A_0-B_0|^2/\Delta$, $|C_{p}(k;\eta)|^2$ can be rewritten as

\be
|C_{p}(k;\eta)|^2 = \frac{\Delta}{Vp^{3-2\Delta}}\frac{|A^*_0-B^*_0|^2}{|A_0-B_0|^2} F[k,\eta] = \frac{\Delta}{Vp^{3-2\Delta}}F[k,\eta] \label{ansat}
\ee The dependence on $\Delta$ is a manifestation of unitarity to leading order; if the integral in  eq.(\ref{ansat}) is calculated over superhorizon modes, then

\be
\sum_{p \lesssim (-1/\eta)}|C_{p}(k;\eta)|^2 =  \frac{ F[k;\eta]\,\Delta}{2\pi^2}~ \int^{(-1/\eta)}_0 \frac{p^2 dp}{p^{3-2\Delta}} =   \frac{F[k;\eta]}{4\pi^2}\,(-1/\eta)^{2\Delta}, \label{inteuni}
\ee  Noting that the $\Delta$ in the numerator in eq.(\ref{ansat}) cancels the single pole in $\Delta$ from the integral giving an $\mathcal{O}(1)$ contribution, which is what is necessary to satisfy the unitarity condition (\ref{unimark}) to leading order in $\Delta$.

This result is similar to that found in the case of particle decay in Minkowski space time\cite{lellomink}: in this case the particles produced from the decay of a parent particle feature a Lorentzian distribution in energy, with width $\Gamma$ the decay width of the parent particle and amplitude $1/\Gamma$, so that the energy integral over the distribution is $\mathcal{O}(1)$. In ref.\cite{lellosup} it is proven to leading order in the perturbative expansion $\mathcal{O}(\Gamma)$ that this narrow distribution of large amplitude is the main reason for the fulfillment of unitarity to leading order in the Wigner-Weisskopf approximation. In the case of de Sitter space time, the distribution function of the particles produced with superhorizon wavevectors is $\propto \Delta /p^{3-2\Delta}$ whose momentum integral over the region of superhorizon momenta is also of $\mathcal{O}(1)$.

Thus in the limit $\Delta \ll 1$ the sum $\sum_{p}\,|C_p(\eta)|^2$ is dominated by the superhorizon momenta and from the unitarity relation (\ref{unimark}) it is found that

\be
\mathrm{Tr} \rho_r(\eta) = |C_k(\eta)|^2 + \sum_p\,|C_p(\eta)|^2 =1  \,.\label{trace1}
\ee To leading order in $\Delta$, the sum is dominated by the superhorizon contributions from both regions of integrations $p \lesssim (-1/\eta)~,~|\vk-\vp|\lesssim (-1/\eta)$ contributing equally, hence

\be
\sum_{p\lesssim (-1/\eta)} |C_{p}(k;\eta)|^2 \simeq \frac{1}{2} \big[ 1- |C_k(\eta)|^2 \big] \,.\label{unitasupho}
\ee Then the factorized form (\ref{ansat}) for superhorizon modes, combined with eqn. (\ref{unitasupho}) leads to

\be
F[k;\eta] = \frac{2\pi^2}{(-\eta)^{-2\Delta}}\,\big[ 1- |C_k(\eta)|^2 \big]\,, \label{fketa}
\ee and for $-k\eta \gg 1$ and $-p\eta \ll 1$ to leading order in $\Delta$, it is found that

\be
|C_{p}(k;\eta)|^2 = \frac{2\pi^2\,\Delta}{V\,p^3\,(-p\eta)^{-2\Delta}}\,\big[ 1- |C_k(\eta)|^2 \big]\,;\label{Cpetasup}
\ee the same result is valid in the region $-k\eta \gg 1$ with $-|\vk-\vp|\eta \ll 1$ by replacing $p \leftrightarrow |\vk-\vp|$.

The long wavelength limit of eq.(\ref{Cpetasup}) requires a careful treatment. Since $|C_p(\eta)|^2=n_p(\eta)$ is the distribution function of particles, for a fixed volume $V$ there is an infrared divergence in the occupation as $p \rightarrow 0$. However, physically the longest allowed wavelength must be determined by the linear size of the quantization volume, this forces an introduction of an infrared cutoff:

\be
p_m = 1/V^{\frac{1}{3}}\,. \label{pmin}
\ee  This treatment is similar to the case of Bose-Einstein condensation where momentum integrals are cut off in the infrared with a typical momentum $p_m \propto L^{-1}$ with $L$ being the typical size of the system. At the end of the calculation of thermodynamic variables one takes $L\rightarrow \infty$ with a careful analysis of the infrared behavior; the remainder of this calculation proceed in much the same manner.

The definition of the lower momentum cutoff $p_m$ may differ from eq.(\ref{pmin}) by overall constants of $\mathcal{O}(1)$; however, as will be shown in detail in the analysis that follows, this proportionality constant would yield an irrelevant contribution in the limit $\Delta \ll 1$.

Now the calculation of the entanglement entropy is straightforward: Consider

\be
I= \sum_{p \leq (-1/\eta)} |C_{p}(k;\eta)|^2 \ln\Big[ |C_{p}(k;\eta)|^2\Big] \equiv I_1 + I_2 \label{Idef}
\ee with

\bea I_1 & = &  \Big[1-|C_k(\eta)|^2\Big]\,\ln\Big[2\pi^2\Delta\,\Big[1-|C_k(\eta)|^2\Big] \Big]\,\Delta \,\int^{(-1/\eta)}_{p_m} (-p\eta)^{2\Delta} \frac{dp}{p} \nonumber \\ & = &  \frac{1}{2}\, \Big[1-|C_k(\eta)|^2\Big]\,\ln\Big[2\pi^2\Delta\,\Big[1-|C_k(\eta)|^2\Big] \Big]\, \,\Big[1-x_m^{2\Delta}\Big]   \label{I1}\eea where the following definition has been made.

\be
x_m = (-p_m\eta) \label{xm}
\ee Evaluating $I_2$ can be done by changing integration variables to $x = -p\eta$ which produces

\bea
I_2 & = & - \Big[1-|C_k(\eta)|^2\Big]\,\Delta  \,\int^{1}_{x_m } x^{2\Delta-1} \ln \Big[\frac{x^{3-2\Delta}}{x^3_m} \Big] \, dx \nonumber \\ & = & \frac{1}{2}\,\Big[1-|C_k(\eta)|^2\Big] \Bigg\{ \frac{3-2\Delta}{2\Delta}\,\Big[1-\left( {x_m} \right)^{2\Delta}\Big] + \\ &
& 3 \ln[x_m] \, \Big[ 1-\left( {x_m} \right)^{2\Delta} \Big] + (3-2\Delta) \Bigg[1-(x_m)^{2\Delta}\Bigg] \ln\left[ {x_m} \right] \Bigg\}\,. \label{I2}
\eea It is now clear that the limit $x_m \rightarrow 0$ may be carried out safely safely in $I_1$ and in the terms that \emph{do not feature poles in $\Delta$} in  $I_2$. The terms in $I_2$ that feature the $\ln[x_m]$ and the (single) pole in $\Delta$, namely $(3/2\Delta) \times [1-(x_m)^{2\Delta}]$   yield the leading contribution for $\Delta, x_m  \ll 1$.

Therefore for $\Delta \ll 1$ and $x_m \ll 1 $, to leading order, the entanglement entropy is found to be

\bea \mathcal{S}(\eta)  & \simeq &  \frac{\alpha}{(k\eta)^2}\,e^{-\frac{\alpha}{(k\eta)^2}}  - \Big[1-e^{-\frac{\alpha}{(k\eta)^2}}\Big]\,\ln\Big[1-e^{-\frac{\alpha}{(k\eta)^2}}\Big] \nonumber \\ & + &
\Big[1-e^{-\frac{\alpha}{(k\eta)^2}}\Big]~ \Bigg\{ \ln\Big[ \frac{1}{2\pi^2\,\Delta}\Big]+\frac{3}{2\Delta}\,\Big[W[\eta] - 1 + e^{- W[\eta]}\Big]+\mathcal{O}(\Delta) \Big]\Bigg\}\label{entrofinDS}\eea where

\be
W[\eta] = \frac{2\Delta}{3}~ \ln\Big[ V_{ph}(\eta)\,H^3 \Big]~~;~~ V_{ph}(\eta) = V\,(C(\eta))^3 \,,\label{weta}
\ee with $C(\eta) = a(t(\eta))$ is the scale factor and $\alpha$ is given in eqn. (\ref{alfadef}).  The function $W[\eta] - 1 + e^{- W[\eta]}$ is   manifestly (semi) positive and monotonically increasing,  behaving as $\simeq W^2/2$ for $W \ll 1 $ and as $\simeq W$ for $W\gg 1$. As $\eta \rightarrow 0$ the entanglement entropy grows monotonically with the physical volume.

 An important consequence of unitarity is that the dependence of the entanglement entropy on the initial conditions is only through $\alpha$.

The logarithmic volume dependence is similar to the result obtained in Minkowski space time, and its interpretation is that asymptotically the entropy saturates to the logarithm of the number of accessible states in phase space, which is proportional to the volume. However in the expanding cosmology it is the physical volume that enters in the final expression; as the cosmological expansion proceeds  the available phase space increases as more and more wavevectors  cross  the Hubble radius. Furthermore the infrared enhancement from light fields during inflation translate in the $\ln[\Delta]$. It is clear from the expression above that the definition of $p_m$ in (\ref{pmin}) differed by a proportionality constant $\mathcal{C} \simeq \mathrm{O}(1)$, the expression above would have been modified by an term $\sim \Delta \ln[\mathcal{C}]\ll 1$ which can be safely neglected, thus confirming that the choice of the minimal value of the momentum (infrared cutoff) (\ref{pmin}) is insensitive to multiplicative factors of $\mathcal{O}(1)$ for $\Delta \ll 1$.

\section{Conclusions and further questions.}

The recent CMB data from Planck distinctly shows a persistence of large scale anomalies,  among them   a suppression of the power spectrum for large scales, in the region of the Sachs-Wolfe plateau for $l \lesssim 10$. Motivated by the possibility that these anomalies, in particular the suppression of power at low multipoles,  is of primordial origin perhaps heralding new physics on superhorizon scales, we studied the effect of initial conditions arising from a rapid evolution of the inflaton during a brief stage prior to slow roll. Such a rapid evolution, or ``fast roll'' stage leads to the equations for the mode functions of scalar and tensor perturbations that features a potential which is localized in conformal time. The effect of this potential translates into non-Bunch-Davies conditions on the mode functions during the slow roll stage, the Bogoliubov coefficients being determined by the properties of the potential during the pre-slow roll stage.

Implementing methods from potential scattering theory we obtained general properties of these Bogoliubov coefficients, in particular their superhorizon and sub-horizon behavior. The effect of these initial conditions on the power spectra of scalar and tensor perturbations are encoded in an initial condition transfer function $\mathcal{T}(k)$. We showed that for wavevectors that exited the Hubble radius during the
very early stages of slow roll the large scale transfer function $\mathcal{T}(k\approx 0)$ leads to a suppression of the power spectrum for attractive potentials, such as those found previously for the case of a ``fast-roll stage''\cite{boyan3,hectordestri,reviunos,lellor}. Furthermore for modes that are inside the Hubble radius during most of the slow roll stage $\mathcal{T}(k) \lesssim 1/k^4$ suggesting that the effect of initial conditions determined by pre-slow roll stage is strongly suppressed for higher multipoles and would not modify the small scale aspects of the CMB, such as acoustic peaks.

Since the initial conditions impact mainly large scales, we were motivated to study their effect on the infrared sector of typical    minimally coupled scalar field theories  with typical self-interactions $\lambda \phi^p$ with $p = 3, 4$ when the slow roll stage is a (nearly) de Sitter cosmology. The correlation functions of light scalar fields with mass $M\ll H$ (H is the Hubble parameter during de Sitter inflation), feature infrared divergences manifest as poles in $\Delta = M^2/3H^2$. These infrared divergences lead to a dynamical generation of mass if the bare mass of the scalar field vanishes.

For $p=3$ we find that the infrared singularity of bare massless theory leads to the formation of a non-perturbative condensate which reaches a fixed value at long times and implies the dynamical generation of a mass $ M = \sqrt{3}\, H   \left( \frac{\lambda}{2 \pi H} \right)^{1/3} \Big[\mathcal{T}(0) \Big]^{1/6} $. For $p=4$ we find $M = H \left[ \frac{9 \lambda\,\mathcal{T}(0)}{2 \pi^2}  \right]^{1/4}$. In both cases the emergence of a dynamical infrared generated mass yields scalar power spectra with \emph{anomalous dimensions} that depend non-analytically on initial conditions, namely
$P_s(k) \propto k^{\Delta}$ where for $p=3,4$ respectively we find
\be  \Delta_{(3)}  =   \Bigg[ \frac{\lambda\,\sqrt{\mathcal{T}(0)}}{2\pi\,H}\Bigg]^\frac{2}{3}\,  ~~;~~ \Delta_{(4)}   =   \Big[\frac{\lambda \, \mathcal{T}(0)}{2\pi^2} \Big]^{\frac{1}{2}} \,.
\ee

In an expanding cosmology all the quanta of a field can decay into quanta of the \emph{same field} as a consequence of the lack of energy conservation and kinematic thresholds. The time dependent decay width of single particle states are enhanced by the infrared divergences that are also responsible for the dynamical generation of mass. We obtain the modification of the decay width for single particle states induced by the non-Bunch -Davies initial conditions, for $p=3$ we find
\be \Gamma(k;\eta) \simeq \frac{9\lambda^2 \mathcal{T}(0)\mathcal{T}^2(k)}{\pi^2H^2\Delta (-\eta)(-k\eta)^6} \ee

The decay of a single particle state yields an \emph{entangled} quantum state of the daughter particles, entanglement being a consequence of momentum conservation. We implement field theoretical version of the Wigner-Weisskopf method adapted to inflationary cosmology to obtain the full quantum state that results from the time evolution and decay of an initial single particle state. This method yields manifestly unitary time evolution of the quantum state. In ref.\cite{lellosup} it was realized that this quantum state features entanglement and correlations between sub and superhorizon quanta. Tracing over the superhorizon degrees of freedom leads to an entanglement entropy that grows as more modes exit the horizon during inflation. We obtain the modifications of this entanglement entropy from non-Bunch-Davies initial conditions. The main change to the entanglement entropy from non-Bunch-Davies initial conditions is through its
dependence   on  the decay width.

In all cases studied in this article, the initial conditions from a ``fast roll'' stage prior to slow roll that result in an initial condition transfer function that suppresses the power of scalar perturbations at large scales, also result in a suppression of the infrared effects: dynamical masses, anomalous dimensions of scalar power spectra and decay widths of quantum states.

\acknowledgements

 L. L. and D.B. acknowledge partial support from NSF-PHY-1202227. R. H.
was supported in part by the Department of Energy under grant DE-FG03-91-ER40682. He
would also like to thank the Cosmology group at UC Davis for hospitality while this work
was in progress.

\appendix

\section{Calculation of $\Sigma(k;\eta_1,\eta_2)$}\label{appA}

In this appendix we calculate the self-energy (\ref{sigma}) to leading order in $\Delta$ and in the long time limit $\eta_1,\eta_2 \rightarrow 0$.

The first step is to perform the angular integration in (\ref{sigma}). Making the substitution $p \equiv |k-q| = \sqrt{k^2+q^2-2kq \cos\theta}$ and $d(\cos\theta) = -p \, dp/kq$ so that

\be
\int d^3q f(|q|)g(|k-q|) = 2\pi \int d(\cos \theta) \int dq ~q^2 [...] =\frac{2 \pi}{k} \int^{\infty}_0 dq~ q ~f(|q|) \int^{k+q}_{|k-q|} dp~ p ~ g(|p|)
\ee This simplifies the integration to

\be
\begin{split}
& \Sigma(k,\eta_1,\eta_2) = \frac{9 \lambda^2}{\pi^2 H^2 k \eta_1 \eta_2}  S^{*}(k, \eta_1) S(k, \eta_2) \int^{\infty}_0 dq \, q S(q, \eta_1) S^{*}(q, \eta_2) \int^{k+q}_{|k-q|} dp \, p S(p, \eta_1) S^{*}(p, \eta_2) \\
& \equiv \frac{9 \lambda^2}{\pi^2 H^2 k \eta_1 \eta_2}  S^{*}(k, \eta_1) S(k, \eta_2) J(k, \eta_1, \eta_2) \label{formula}
\end{split}
\ee where

\be
J(k,\eta_1,\eta_2) = \int^{\infty}_0 dq \,q \, S(q, \eta_1) S^{*}(q, \eta_2)\, \int^{k+q}_{|k-q|} dp \,p \,S(p, \eta_1) S^{*}(p, \eta_2) \label{nastyintegral}
\ee

As with the tadpole, this integral features infrared divergences for massless, minimally coupled fields. From the discussion of the tadpoles, it should be clear that there are infrared divergences for $q,p \rightarrow 0 $, namely in the integration regions $q \simeq 0~;~q \simeq k$. The integral is evaluated with the same method as for the tadpole,  isolating the regions of infrared divergences by introducing an infrared cutoff, keeping the most infrared singular terms of the mode functions in the band of wavevectors up to the infrared cutoff extracting the leading order poles in $\Delta$  and set $\nu = 3/2$    for wavevectors larger than the cutoff since these integrals are infrared finite for finite cutoff in the limit $\Delta \rightarrow 0$.    Therefore, we write  in obvious notation

\be
J = \int^{\mu}_0 dq \, [...] + \int^{\infty}_{\mu} dq \, [...]  \equiv J_{<} + J_{>} \label{jsplit}
\ee

The $J_{<}$ integral is evaluated by using $q < \mu \sim 0$ so that with $k\gg \mu$ the argument  of the $p$-integral can be evaluated at $p=k$ and the $p$ integral becomes simply $ 2k q S(k, \eta_1) S^{*}(k, \eta_2)$ and

\be
\begin{split}
& J_{<}= \int^{\mu}_{0} dq \, q \, S(q, \eta_1) S^{*}(q, \eta_2)\, \int^{k+q}_{|k-q|} dp \,p \,S(p, \eta_1) S^{*}(p, \eta_2) \\
& \sim 2k \, S(k, \eta_1) \, S^{*}(k, \eta_2) \int^{\mu}_{0} dq \, q^2 \, S(q, \eta_1) S^{*}(q, \eta_2)
\end{split}
\ee Using the long wavelength and long time form of the mode functions given by eqn. (\ref{modlwlate}) we find

\be
J_{<}= S(k, \eta_1) \, S^{*}(k, \eta_2) \left[ \left(\frac{4}{\eta_1 \eta_2}\right)^{\nu-1/2} \frac{k \, \Gamma^2(\nu)}{\pi} \right] \,\mathcal{T}(0) \,\frac{\mu^{2\Delta}}{2\Delta}
\ee

To evaluate the $J_{>}$ integral, care must be taken around the poles. There will be infrared divergences for $q=k$ so that the integral is separated as

\be
J_{>} = \int^{\infty}_{\mu} dq [ ...] = \underbrace{\int^{k-\mu}_{\mu} dq [ ...]}_{J^{(a)}_{>}} +  \underbrace{\int^{k}_{k-\mu} dq [ ...]}_{J^{(b)}_{>}} + \underbrace{\int^{k+\mu}_{k} dq [ ...]}_{J^{(c)}_{>}} +   \underbrace{\int^{\infty}_{k+\mu} dq [ ...]}_{J^{(d)}_{>}}
\ee Since the integrals away from the infrared limit, namely $J^{(a)/(d)}_{>}$, are finite for finite $\mu$, we can set in these integrals $\nu=3/2$ as they do not feature poles in $\Delta$. In which case, these integrals are subleading with respect to $\Delta$ and need not be considered for a leading order calculation.

The only integrals remaining for the leading order contribution are $J^{(b,c)}_{>}$.  Consider
\be J^{(b)}_{>} = \int^{k}_{k-\mu} dq  \,q \, S(q, \eta_1) S^{*}(q, \eta_2)\, \int^{k+q}_{|k-q|} dp \,p \,S(p, \eta_1) S^{*}(p, \eta_2)\,, \label{jlessb}\ee after the change of variable $q=k-r$, to leading order we obtain
\be J^{(b)}_{>}   \simeq  k \, S(k, \eta_1) \, S^{*}(k, \eta_2) \int^{\mu}_{0} dr    \int^{2k+r}_{r} dp \,p \,S(p, \eta_1) S^{*}(p, \eta_2)\,, \label{jless2}\ee the leading order contribution arises from the lower limit of the $r$ integral, this contribution is obtained by integrating in a small region around the lower limit using the mode functions (\ref{modlwlate}) and approximating $\mathcal{T}(p) \simeq \mathcal{T}(0) + \mathcal{O}(p^2)+\cdots$ and keeping only  the $p=0$ term in this expansion because the higher order terms will not yield poles in $\Delta$, we find
\be \int^{2k+r}_{r} dp \,p \,S(p, \eta_1) S^{*}(p, \eta_2) =  \frac{\Gamma(\nu) \Gamma(\nu-1)}{2\pi^2} \left(\frac{4}{\eta_1 \eta_2} \right)^{\nu} \,  \frac{\pi}{4} \Big(\eta_1\,\eta_2\Big)^{1/2}\, \mathcal{T}(0)~ r^{2-2\nu} +\cdots
\ee where the dots stand for terms that will not yield poles in $\Delta$ as $\Delta \rightarrow 0$. Finally carrying out the $r$-integral we find
\be J^{(b)}_{>}   = k \, S(k, \eta_1) \, S^{*}(k, \eta_2) \frac{\Gamma(\nu) \Gamma(\nu-1)}{2\pi^2} \left(\frac{4}{\eta_1 \eta_2} \right)^{\nu} \, \frac{\pi}{4} \Big(\eta_1\,\eta_2\Big)^{1/2} \frac{\mu^{2\Delta}}{2\Delta} + \cdots \,. \label{jbgreatfin} \ee

 The next term $J^{(c)}_>$   can be evaluated in a similar manner, but now changing variables in the $q$-integral to $q= k+r$  and recognizing that the lower limit in the $p$-integral is now $q-k =r$ upon changing variables in the $q$-integral. Again the $p$ integral is dominated by the lower limit which can be extracted just as in the previous case finally leading to
 \be J^{(c)}_{>}= J^{(b)}_{>} \,. \label{eqterms}\ee
 Expanding the pole terms
 \be \frac{\mu^{2\Delta}}{2\Delta} = \frac{1}{2\Delta} + \ln[\mu] + \cdots \,\label{logs}\ee all the terms with $\ln[\mu]$ will cancel among all the different contributions, this is easily seen by taking the $\mu$ derivative of $J$ given by eqn. (\ref{jsplit}) as the arbitrary cutoff $\mu$ has been introduced simply to split  the integrals and the total integral cannot depend on $\mu$.

Finally, to leading order

\be
 J = J_< + J^{(b)}_> + J^{(c)}_> + \mathcal{O}(\Delta^0) = 2k \frac{S(k,\eta_1) S^*(k, \eta_2)}{\eta_1 \eta_2} \mathcal{T}(0)~ \Big(\frac{1}{\Delta} +   \mathcal{O}(\Delta^0) \Big)  \\
 \label{pole}
\ee Combining this result with (\ref{formula}) we finally find,
\be
\Sigma(k,\eta_1,\eta_2) = \frac{18\,\lambda^2\,\mathcal{T}(0)}{\pi^2\,H^2\,\Delta   } ~  \frac{\left| S(k, \eta_1)\right|^2}{(\eta_1)^2}\frac{ \left| S(k, \eta_2)\right|^2}{(  \eta_2)^2}\,    + \mathcal{O}(\Delta^0) \,. \ee

\section{Wigner-Weisskopf theory and unitarity} \label{appb}
In this apprendix we summarize the main aspects of the non-perturbative Wigner-Weisskopf method to study the quantum state from particle decay for consistency. More details are available in refs.\cite{boyholds,boyquasi,lellosup,lellomink}.

The interaction picture   states are expanded in terms of  Fock states associated with the creation and annihilation operators $\alpha_k,\alpha^\dagger_k$, namely

\be
| \Psi(\eta) \rangle_I = \sum_{n} C_n (\eta) |n\rangle
\ee As shown in earlier, the time evolution of a state in the interaction picture is given by

\be
i \frac{d}{d\eta} | \Psi(\eta) \rangle_I = \hat{H}_I(\eta) | \Psi(\eta) \rangle_I
\ee so that the (conformal) time evolution of the coefficients is given by

\be
\frac{d}{d\eta} C_n(\eta) = -i \sum_{m} C_m(\eta) \langle n | \hat{H}_I(\eta) | m \rangle
\ee While this is exact, the solution is an \emph{infinite hierarchy} and finding an exact solution is impractical. This can be vastly simplified by making the assumption that the initial state, $|A\rangle$, only couples to a single set of intermediate states, $|\kappa\rangle$, where this assumption is exact if the situation is confined to processes of $\mathcal{O}(H_I)$ (which is valid for this work). Under this assumption, the coefficients obey

\be
\begin{split}
& \frac{d}{d\eta} C_{A}(\eta) = -i \sum_{\kappa} \langle A | H_{I}(\eta) | \kappa \rangle C_{\kappa}(\eta) \\
& \frac{d}{dt} C_{\kappa}(\eta) = -i \langle \kappa | H_{I}(\eta) | A \rangle C_{A}(\eta)
\end{split} \label{coeff}
\ee where $\sum_{\kappa}$ is over all states that couple to $|A\rangle$ via first order in $H_I$.

Considering the general situation of particle decay, $A \rightarrow \kappa_1, \kappa_2, ...$, where initally at some time, $\eta = \eta_o$, the state is given by $|\Psi (\eta_o) \rangle = |A\rangle$. This is equivalent to the initial condition $C_n(\eta_o) = \delta_{n,A}$. Upon integrating the second of \ref{coeff}, one obtains

\be
\begin{split}
& C_{\kappa}(\eta) = -i \int_{0}^{\eta} d\eta' \langle \kappa | H_{I}(\eta') | A \rangle C_{A}(\eta') \\
& \frac{d}{d\eta} C_{A}(\eta) = - \int_{0}^{\eta} d\eta' \sum_{\kappa} \langle A | H_{I}(\eta) | \kappa \rangle \langle \kappa | H_{I}(\eta') | A \rangle C_{A}(\eta') \label{wwcoeff}
\end{split}
\ee It proves useful to make the definition

\be
\Sigma_{A}(\eta, \eta') = \sum_{\kappa} \langle A | H_{I}(\eta) | \kappa \rangle \langle \kappa | H_{I}(\eta') | A \rangle
\ee Note that this is equal to \ref{sigma}. Then

\be
\frac{d}{d\eta} C_{A}(\eta) = - \int_{\eta_o}^{\eta} d\eta' \Sigma_{A}(\eta,\eta') C_{A}(\eta') \label{coeffa}
\ee

The relation between this method and the Dyson resummation is discussed in detail in ref. \cite{boyan}. It can be shown that this treatment is non-perturbative and the time evolution of the coefficients are slow which justifies a derivative expansion. The derivative expansion is done by introducing the term

\be
W_0(\eta,\eta') = \int^{\eta'}_{\eta_o} d\eta'' \Sigma_{A}(\eta,\eta'') ~~;~~ \frac{d}{d\eta'} W_0(\eta,\eta') = \Sigma_{A}(\eta,\eta') ~~;~~  W_0(\eta,\eta_o)=0
\ee So that integrating \ref{coeffa} by parts leads to

\be
\int_{\eta_o}^{\eta} d\eta' \Sigma_{A}(\eta,\eta') C_{A}(\eta') = W_0(\eta,\eta)C_{A}(\eta)-\int_{\eta_o}^{\eta} d\eta' W_0(\eta,\eta') \frac{d}{d\eta'}C_{A}(\eta')
\ee For a weakly interacting theory, such that $H_I \sim \mathcal{O}(\lambda)$ and $\lambda \ll 1$, the second term is at higher order in perturbation theory and may be discarded. To leading order, \ref{coeffa} simplifies drastically to

\be
\frac{d}{d\eta} C_{A}(\eta) = -  W_0(\eta,\eta)C_{A}(\eta) + \mathcal{O}(\lambda^4)
\ee with the simple solution

\be
C_A(\eta) = e^{-\int^{\eta}_{\eta_o} d\eta' W_0(\eta',\eta')} \label{initialcoeff}
\ee

Interpretation of this result follows from the analysis in Minkowski spacetime. It has been shown that the imaginary part of the integral will provide the second order energy shift while the real part provides the decay width, similar to Fermi's golden rule. This is made explict in the literature with the result that

\be
\int^{\eta'}_{\eta_o} d\eta'' \Sigma_{A}(\eta',\eta'') = i \delta E^{(1)}_A (\eta') + \frac{1}{2} \Gamma(\eta')
\ee Where the real part matches \ref{decay} exactly. Finally, the full time dependence of the coefficient can be written as

\be
C_A (\eta) = e^{-i \int d\eta' \delta E^{(1)} (\eta')} e^{-\frac{1}{2} \int d\eta' \Gamma_A (\eta')}
\ee Since the probability of measuring particle $A$ is $|C_A|^2$ and with the discussion in section \ref{decaySec}, the interpretation of $\Gamma$ as the decay rate is clear. It has also been shown that the Wigner Weisskopf method produces the same results for the self consistent mass generation discussed earlier \cite{boyan}.

One the main goals is to study the entanglement entropy from tracing over superhorizon degrees of freedom. Thus
it is important to make sure that the loss of information encoded in the entanglement entropy is a genuine effect of the tracing procedure and not a consequence of approximations in the evolution of the quantum state. In this appendix, the discussion follows ref. \cite{lellomink,lellosup} where it is shown that the Wigner-Weisskopf approximation and its Markovian implementation maintain unitary time evolution.

Using (\ref{wwcoeff}) consider

\be
\sum_{\kappa}|C_\kappa(\eta)|^2 = \int_{\eta_0}^{\eta}d\eta_1 C^*_A(\eta_1)\int_{\eta_0}^{\eta}d\eta_2\Sigma(\eta_1,\eta_2) C_A(\eta_2). \label{sumkapa}
\ee Inserting $1=\Theta(\eta_1-\eta_2)+\Theta(\eta_2-\eta_1)$,  it follows that

\bea
\sum_{\kappa}|C_\kappa(\eta)|^2 & = &  \int_{\eta_0}^{\eta}d\eta_1 C^*_A(\eta_1)\int_{\eta_0}^{\eta_1}d\eta_2\Sigma(\eta_1,\eta_2) C_A(\eta_2)  \nonumber \\ & + & \int_{\eta_0}^{\eta}d\eta_2 C_A(\eta_2)\int_{\eta_0}^{\eta_2}d\eta_1\Sigma(\eta_1,\eta_2) C^*_A(\eta_1).\label{shuffle}
\eea Using $\Sigma (\eta_1,\eta_2) = \Sigma^*(\eta_2,\eta_1)$, relabelling $\eta_1 \leftrightarrow \eta_2$ in the second line of (\ref{shuffle}) and using (\ref{coeffa}), one can show

\bea
\sum_{\kappa}|C_\kappa(\eta)|^2  &=& - \int_{\eta_0}^{\eta}d\eta_1 \Big[ C^*_A(\eta_1)\frac{d}{d \eta_1}C_A(\eta_1) + C_A(\eta_1) \frac{d}{d \eta_1}C^*_A(\eta_1)\Big]\nonumber\\
&=& - \int_{\eta_0}^{\eta}d\eta_1 \frac{d}{d\eta_1} |C_A(\eta_1)|^2 = 1-|C_A(\eta)|^2 \label{unita}
\eea where the initial condition $C_A(\eta_0)=1$ has been used. This is the statement of unitary time evolution, namely

\be
|C_A(\eta)|^2 + \sum_{\kappa}|C_\kappa(\eta)|^2 = |C_A(\eta_0)|^2 \label{unitime}
\ee

To leading order in the Markovian approximation, the unitarity relation becomes

\be
\sum_{\kappa}|C_\kappa(\eta)|^2   =   -2 \int_{\eta_0}^{\eta}  \Big|C_A(\eta_1)\Big|^2 \,\mathrm{Re}\Big[W_0(\eta_1,\eta_1)\Big] \,d\eta_1  =  1-|C_A(\eta)|^2 \label{unimark}
\ee where $C_A(\eta_0)=1$.


\begin{thebibliography}{999}

\bibitem{staro2} A. A. Starobinsky, JETP Lett. 30, 682 (1980); Phys. Lett. \textbf{91B}, 99 (1980);
 V. F. Mukhanov, G. V. Chibisov, Soviet Phys. JETP Lett. 33, 532 (1981). % quad sup 6

 \bibitem{guth} A. H. Guth, Phys. Rev. \textbf{D23}, 347 (1981).

 \bibitem{linde} A. A. Linde, Phys. Lett. \textbf{108B}, 389 (1982); Phys. Lett. \textbf{116B},335 (1982);
 Phys. Lett. \textbf{129B},177 (1983).

 \bibitem{al} A. A. Albrecht and P. Steinhardt, Phys. Rev. Lett. \textbf{48}, 1220 (1982).





\bibitem{mukh} V. F. Mukhanov, H. A. Feldman , R. H. Brandenberger, Phys. Rept. 215, 203 (1992). % quad sup


\bibitem{riotto2} A. Riotto, arXiv: hep-ph/0210162. % quad sup 9

\bibitem{baumann} D. Baumann,  arXiv:0907.5424.

\bibitem{giov} M. Giovannini, Int. J. Mod. Phys. D14 363 (2005). % quad sup 10

\bibitem{kolb} J. Lidsey, A. R. Liddle, E. Kolb, Rev. of Mod. Phys. \textbf{69}, 373 (1997).



\bibitem{wmap7} E. Komatsu \emph{et.al.} (WMAP collaboration),  	Astrophys.J.Suppl.\textbf{192}, 18 (2011).
\bibitem{wmap9} G. Hinshaw \emph{et.al.} (WMAP collaboration),  	 Astrophys.J.Suppl. \textbf{208}, 19.

\bibitem{planck}  P. A. R. Ade \emph{et.al.} (PLANCK collaboration),  arXiv:1303.5076;  arXiv:1303.5082;
arXiv:1303.5075; arXiv:1303.5083.


\bibitem{cobe}G.  Hinshaw, A. J.  Branday, C. L.  Bennett,   \emph{et al.} , ApJ, \textbf{464},
L25 (1996).

\bibitem{bondlow} Bond, J. R., Jaffe, A. H.,  Knox, L.   Phys. Rev. \textbf{D 57}, 2117 (1998).

\bibitem{wmaplow} D. N. Spergel,   \textbf{et al.} (WMAP collaboration),  ApJS, \textbf{148}, 175 (2003).
\bibitem{berera1} A. Berera, L.-Z. Fang, G. Hinshaw,  	Phys.Rev. \textbf{D57}, 2207 (1998).

\bibitem{berera2}   A. Berera, A. F. Heavens,  	Phys.Rev.\textbf{D62} 123513 (2000).



\bibitem{francis} C. L. Francis and J. A. Peacock, MNRAS\textbf{ 406}, 14 (2010).

\bibitem{rassat} A. Rassat, J.-L. Starck, F.-X. Dupe,  	arXiv:1303.4727.

\bibitem{grup} A. Gruppuso, P. Natoli, F. Paci, F. Finelli, D. Molinari, A. De Rosa, N. Mandolesi,  	arXiv:1304.5493.

\bibitem{copi} C. J. Copi, D. Huterer, D. J. Schwarz, G. D. Starkman, Advances in Astronomy vol. 2010, Article ID 847541 (2010); C. J. Copi, D. Huterer, D. J. Schwarz, G. D. Starkman, arXiv:1310.3831; A. Yoho, C. J. Copi, G. D. Starkman, A. Kosowsky,  arXiv:1310.7603.

    \bibitem{bunch} T. S. Bunch and P. C. Davies, Proc. R. Soc. A360, 117 (1978); N. D. Birrell and P. C. W. Davies, Quantum ﬁelds in curved space, (Cambridge Monographs in Mathematical Physics, Cambridge University Press, Cambridge, 1982). % quad sup 19

 \bibitem{ini1} N. Kaloper, M. Kleban, A. Lawrence, S. Shenker and L. Susskind, JHEP \textbf{0211}, 037
(2002).

\bibitem{ini2} B. Greene, K. Schalm, J. P. van der Schaar and G. Shiu, In the Proceedings of 22nd
Texas Symposium on Relativistic Astrophysics at Stanford University, Stanford,
California, 13-17 Dec 2004, pp 0001 [arXiv:astro-ph/0503458].

\bibitem{ini3}
R. Easther, W. H. Kinney and H. Peiris, JCAP\textbf{ 0508}, 001 (2005).

\bibitem{ini4} R. Brunetti, K. Fredenhagen and S. Hollands, JHEP \textbf{0505}, 063 (2005).

\bibitem{ini5} K. Goldstein and D. A. Lowe, Nucl. Phys.\textbf{ B 669}, 325 (2003).
\bibitem{holini1} H. Collins and R. Holman, Phys. Rev. \textbf{D 70}, 084019 (2004).
\bibitem{holini2} H. Collins, R. Holman and M. R. Martin, Phys. Rev. \textbf{D 68}, 124012 (2003);
C. P. Burgess, J. M. Cline, F. Lemieux and R. Holman,
JHEP 0302, 048 (2003); C. P. Burgess,
J. M. Cline and R. Holman, JCAP 0310, 004 (2003).


\bibitem{martin} J. Martin and R. Brandenberger, Phys. Rev. D 68, 063513; R. H. Brandenberger
and J. Martin, Int. J. Mod. Phys. A 17, 3663 (2002).

\bibitem{daniels} U. H. Danielsson, Phys. Rev. D 66, 023511 (2002);
U. H. Danielsson, JHEP 0207, 040 (2002).

\bibitem{picon} C. Armendariz-Picon, JCAP 0702, 031 (2007).


\bibitem{holtol1} R. Holman, Andrew J. Tolley,  JCAP \textbf{0805}, 001 (2008).

\bibitem{holtol2} Nishant Agarwal, R. Holman, Andrew J. Tolley, Jennifer Lin,   JHEP \textbf{1305}, 085 (2013).

\bibitem{ganc} J. Ganc,  Phys. Rev.\textbf{ D 84}, 063514 (2011).

\bibitem{parker1} I. Agullo, J. Navarro-Salas, L. Parker,  JCAP \textbf{1205}, 019 (2012).

\bibitem{parker2} I. Agullo,  L. Parker,    Phys.Rev.\textbf{D83}, 063526 (2011).  	

\bibitem{porto} R. Flauger, D. Green, R. A. Porto, arXiv:1303.1430.

\bibitem{dustin} A. Aravind, D. Lorshbough, S. Paban,  JHEP \textbf{1307}, 076 (2013).

\bibitem{ganckoma} J. Ganc, E. Komatsu, Phys. Rev. \textbf{D86}, 023518 (2012).

\bibitem{lindefast} A. Linde, JHEP \textbf{11},  052 (2001).


        \bibitem{contaldi} C. Contaldi, M. Peloso, L. Kofman, A. Linde,  	JCAP \textbf{0307}, 002 (2003).

        \bibitem{boyan3} D. Boyanovsky, H. J. de Vega, N. G. Sanchez, Phys. Rev. \textbf{D74}, 123006, 123007 (2006).

\bibitem{hectordestri}  C. Destri, H. J. de Vega, N. G. Sanchez, Phys.Rev.\textbf{D81}063520 (2010); C. Destri, H. J. de Vega, N. G. Sanchez,  Phys.Rev.\textbf{D78}, 023013 (2008); F. J. Cao, H. J. de Vega, N. G. Sanchez, Phys.Rev.\textbf{D78}, 083508 (2008).

    \bibitem{reviunos} D. Boyanovsky, C. Destri, H. J. de Vega, N. G. Sanchez,  Int.J.Mod.Phys.\textbf{A24}, 3669 (2009).

    \bibitem{lasenby} W.J. Handley, S.D. Brechet, A.N. Lasenby, M.P. Hobson, arXiv:1401.2253.

    \bibitem{lellor} L. Lello, D. Boyanovsky,  arXiv:1312.4251.

\bibitem{parkglenz} M. M. Glenz, L. Parker, Phys.Rev.\textbf{D80}, 063534 (2009); M. M. Glenz,  arXiv:0905.2641.

    \bibitem{schwarz} E. Ramirez, D. J. Schwarz,  	Phys.Rev.\textbf{D80}, 023525 (2009);  	Phys. Rev. \textbf{D 85}, 103516 (2012).

    \bibitem{amjad} A. Ashoorioon, K. Dimopoulos, M. M. Sheikh-Jabbari, G. Shiu,  	arXiv:1306.4914; A. Ashoorioon, G. Shiu,  	JCAP \textbf{1103}, 025 (2011); A. Ashoorioon, A. Krause, arXiV: hep-th/0607001; A. Ashoorioon, A. Krause, K. Turzynski,  	JCAP \textbf{0902}, 014 (2009).

\bibitem{kundu} S. Kundu,
 JCAP {\bf 1202}, 005 (2012).

 \bibitem{jain}  R. K. Jain, P. Chingangbam, J.-O. Gong, L. Sriramkumar and T. Souradeep,
 JCAP \textbf{0901}, 009 (2009);
R. K. Jain, P. Chingangbam, L. Sriramkumar and T. Souradeep,
Phys. Rev. \textbf{D 82}, 023509 (2010).



    \bibitem{weinberg} S. Weinberg, Phys. Rev.\textbf{D72}, 043514 (2005); Phys. Rev.
\textbf{D74}, 023508 (2006).

 \bibitem{seery} D. Seery, Class. Quant. Grav. \textbf{27}, 124005 (2010); JCAP \textbf{0905}, 021 (2009); JCAP \textbf{0802}, 006 (2008); JCAP \textbf{0711}, 025 (2007).

 \bibitem{branrecent} W. Xue, X. Gao, R. Brandenberger, JCAP \textbf{1206}, 035 (2012).



 \bibitem{giddins} S. B. Giddings, M. S. Sloth, JCAP \textbf{1101}, 023 (2011).

 \bibitem{hebe} C. T. Byrnes, M. Gerstenlauer, A. Hebecker, S. Nurmi, G. Tasinato,  	JCAP \textbf{1008}, 006  (2010); M. Gerstenlauer, A. Hebecker, G. Tasinato, JCAP \textbf{1106}, 021 (2011).


\bibitem{bran} W. Xue, K. Dasgupta, R. Brandenberger,  	Phys.Rev.\textbf{D83}, 083520 (2011).

  \bibitem{woodard} S. P. Miao, N. C. Tsamis, R. P. Woodard,  	J.Math.Phys.\textbf{51}, 072503 (2010); R. P. Woodard, arXiv:astro-ph/0310757; T. M. Janssen, S. P. Miao, T. Prokopec, R. P. Woodard,  Class.Quant.Grav.\textbf{25}, 245013 (2008); N. C. Tsamis and R. P. Woodard,   Phys. Lett. \textbf{B
301}, 351 (1993) 351; N. C. Tsamis and R. P. Woodard,   Annals
Phys. \textbf{238},1  (1995); N. C. Tsamis and R. P. Woodard,   Phys.
Rev. \textbf{D 78}, 028501 (2008).

\bibitem{rajaraman} A. Rajaraman,  Phys.Rev.\textbf{D82}, 123522 (2010).

 \bibitem{holmanburgess}  C.P. Burgess, R. Holman, L. Leblond, S. Shandera JCAP \textbf{1003}, 033 (2010);  JCAP \textbf{1010}, 017 (2010).

  \bibitem{riottosloth} A. Riotto and M. S. Sloth, JCAP \textbf{0804}, 030 (2008).

 \bibitem{enq} K. Enqvist, S. Nurmi, D. Podolsky, G. I. Rigopoulos, JCAP \textbf{0804}, 025 (2008).


  \bibitem{boyholds}  D. Boyanovsky, R. Holman, JHEP,Volume \textbf{2011}, Number 5, 47 (2011).

   \bibitem{boyquasi} D. Boyanovsky,  Phys. Rev. D 85, 123525 (2012);  Phys. Rev. D 86, 023509 (2012).

    \bibitem{serreau1} J. Serreau, R. Parentani,  Phys.Rev. \textbf{D87}, 085012 (2013); R. Parentani, J. Serreau,  Phys.Rev. \textbf{D87}, 045020 (2013)   ; F. Gautier, J. Serreau,  arXiv:1305.5705.

     \bibitem{akhmedov} E. T. Akhmedov, A. Roura, A. Sadofyev, Phys.~Rev.\textbf{D82}, 044035 (2010);  E. T. Akhmedov, P. V. Buividovich, Phys.~Rev.\textbf{D78}, 104005 (2008);  E. T. Akhmedov, Mod.Phys.Lett.\textbf{A25},2815 (2010); E. T. Akhmedov, P. V. Buividovich, D. A. Singleton,Phys.Atom.Nucl. \textbf{75} (2012) 525; E. T. Akhmedov, JHEP \textbf{1201}, 066 (2012); E. T. Akhmedov, Ph. Burda  Phys.Rev. \textbf{D86} (2012) 044031; E. T. Akhmedov, Phys.Rev. D87 (2013)
         044049.





\bibitem{staro1} A. A. Starobinski, J. Yokoyama, Phys. Rev. D50, 6357 (1994). % quasi 29 selfconsmass.

\bibitem{richard} R. P. Woodard, J.Phys.Conf.Ser.\textbf{68}, 012032 (2007); S.-P. Miao, R. P. Woodard ;  Phys.Rev.\textbf{D74}, 044019 (2006); R. P. Woodard, arXiv:astro-ph/0502556; T. Brunier, V. K. Onemli, R. P. Woodard, Class.Quant.Grav.\textbf{22}, 59 (2005); E. O. Kahya and V. K. Onemli,
  Phys. Rev. \textbf{D76}, 043512  (2007);  T. Prokopec, O. Tornkvist, R. Woodard, Phys.Rev.Lett.\textbf{89}, 101301 (2002).


 \bibitem{rigo} B. Garbrecht, G. Rigopoulos, Phys. Rev. \textbf{D 84}, 063516 (2011).

\bibitem{garb} B. Garbrecht, T. Prokopec; Phys.Rev. \textbf{D73}  064036  (2006).

\bibitem{arai} T. Arai,  	Class. Quantum Grav. \textbf{29}, 215014 (2012).

 \bibitem{serreau2} J. Serreau,  Phys.Rev.Lett. \textbf{107},  191103    (2011).


\bibitem{boyprem} D. Boyanovsky, R. Holman, S. Prem Kumar,   Phys.~Rev. \textbf{D56}, 1958   (1997).

 \bibitem{boyan}  D. Boyanovsky, H. J. de Vega, Phys.~Rev. \textbf{D70},  063508  (2004);  D. Boyanovsky, H. J. de Vega, N. G. Sanchez,  Phys.~Rev.\textbf{D71} 023509 (2005); Nucl.~Phys. \textbf{B747}, 25 (2006).


\bibitem{moschella} J. Bros, H. Epstein, M. Gaudin, U. Moschella and V. Pasquier, Commun. Math. Phys. \textbf{295}, 261 (2010);   J. Bros, H. Epstein and U. Moschella, arXiv:0812.3513;
J. Bros, H. Epstein and U. Moschella, JCAP \textbf{0802}, 003 (2008).

   \bibitem{leblond} D. P. Jatkar, L. Leblond, A. Rajaraman, Phys.Rev. \textbf{D85}, 024047 (2012).


\bibitem{marolf} D. Marolf, I. A. Morrison, M. Srednicki,  	Class. Quant. Grav. \textbf{30}, 155023 (2013).

\bibitem{lellomink} L. Lello, D. Boyanovsky, R. Holman,   JHEP \textbf{2013},116 (2013).

\bibitem{lellosup} L. Lello, D. Boyanovsky, R. Holman,  arXiv:1305.2441.


\bibitem{smit} M. van der Meulen, J. Smit, JCAP 0711, 023 (2007)











%%%%%%%%%






\end{thebibliography}
\end{document}